\documentclass[acmsmall]{acmart}
\usepackage{hyperref}
\usepackage{latexsym}
\usepackage{mathrsfs}
\usepackage{multirow}
\usepackage{amsmath}
\usepackage{amsfonts}
\usepackage{subfigure}
\usepackage{graphicx}
\usepackage{tabularx}
\usepackage[boxed,ruled,lined]{algorithm2e}
\usepackage{algorithmic}
\usepackage{comment}
\usepackage{balance}
\usepackage{bm}
\newtheorem{remark}{Remark}

\usepackage{soul}

\newtheorem{theorem}{Theorem}
\newtheorem{lemma}{Lemma}

\AtBeginDocument{%
  \providecommand\BibTeX{{%
    \normalfont B\kern-0.5em{\scshape i\kern-0.25em b}\kern-0.8em\TeX}}}
\DeclareMathOperator*{\argmin}{arg\,min}
\DeclareMathOperator*{\argmax}{arg\,max}

\setcopyright{acmlicensed}
\acmJournal{TOIS}
\acmYear{2024} \acmVolume{1} \acmNumber{1} \acmArticle{1} \acmMonth{1}\acmDOI{10.1145/3695867}

\begin{document}
\title{LTP-MMF: Towards Long-term Provider Max-min Fairness \\Under Recommendation Feedback Loops}

\author{Chen Xu}
\affiliation{%
  \institution{Gaoling School of Artificial Intelligence}
    \country{Renmin University of China}
    \\xc\_chen@ruc.edu.cn
}

\author{Xiaopeng Ye}
\affiliation{%
  \institution{Gaoling School of Artificial Intelligence}
   \country{Renmin University of China}
  \\	xpye@ruc.edu.cn
}

\author{Jun Xu}
\authornote{Jun Xu is the corresponding author.}
\affiliation{%
  \institution{Gaoling School of Artificial Intelligence}
    \country{Renmin University of China}
  \\junxu@ruc.edu.cn
}

\author{Xiao Zhang}
\affiliation{%
  \institution{Gaoling School of Artificial Intelligence}
   \country{Renmin University of China}
  \\zhangx89@ruc.edu.cn
}

\author{Weiran Shen}
\affiliation{%
  \institution{Gaoling School of Artificial Intelligence}
    \country{Renmin University of China}
  \\shenweiran@ruc.edu.cn
}

\author{Ji-Rong Wen}
\affiliation{%
  \institution{Gaoling School of Artificial Intelligence}
  \country{Renmin University of China}
  \\jrwen@ruc.edu.cn
}

\begin{abstract}
Multi-stakeholder recommender systems involve various roles, such as users, and providers. Previous work pointed out that max-min fairness (MMF) is a better metric to support weak providers. However, when considering MMF, the features or parameters of these roles vary over time, how to ensure long-term provider MMF has become a significant challenge. We observed that recommendation feedback loops (named RFL) will influence the provider MMF greatly in the long term. 
RFL means that recommender systems can only receive feedback on exposed items from users and update recommender models incrementally based on this feedback. When utilizing the feedback, the recommender model will regard the unexposed items as negative. In this way, the tail provider will not get the opportunity to be exposed, and its items will always be considered negative samples. Such phenomena will become more and more serious in RFL. To alleviate the problem, this paper proposes an online ranking model named Long-Term Provider Max-min Fairness (named LTP-MMF). Theoretical analysis shows that the long-term regret of LTP-MMF enjoys a sub-linear bound. Experimental results on three public recommendation benchmarks demonstrated that LTP-MMF can outperform the baselines in the long term.
\end{abstract}



\maketitle
 
\section{Introduction}\label{sec:intro}

In multi-stakeholder recommender systems (RS), different roles are involved, including users, providers, recommender models, etc~\cite{abdollahpouri2020multistakeholder}. In recent years, provider fairness has received increasing attention, including provider demographic parity fairness~\cite{akpinar2022long,singh2018fairness,Fairco,ge2021towards}, proportion fairness~\cite{cpfair,wu2021tfrom} and max-min fairness~\cite{nips21welf, P-MMF}. 
Max-min fairness (MMF) aims to support worst-off providers and it is proposed based on distributive justice concept~\cite{P-MMF, sep-justice-distributive}. Worse-off providers, who occupy the majority of the platform, cannot survive with the necessary support. Supporting the weak providers will increase the stability of the recommender market and make a good ecosystem~\cite{fairrec, Erickson18}. In this paper, we focused on the amortized max-min fairness~\cite{P-MMF, biega2018equity, nips21welf, balseiro2021regularized}, where fairness accumulated across a series of rankings. 

\begin{figure}[t]  
    
    \centering    
    \subfigure[Feedback loops.]
    {
        \includegraphics[width=0.35\linewidth]{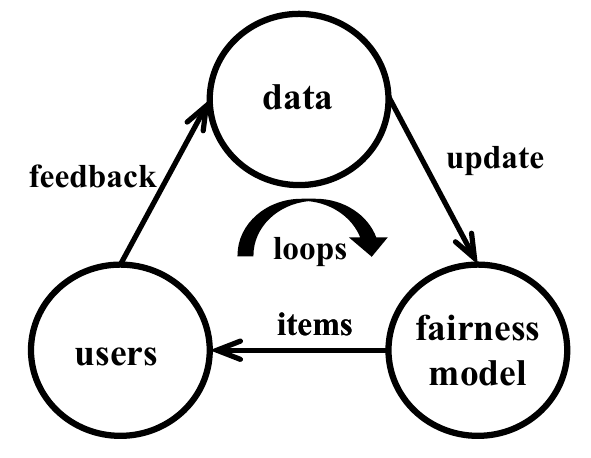}
        \label{fig:intro_loop}
    }
    \subfigure[Simulation of feedback loops.]
    {
        \includegraphics[width=0.35\linewidth]{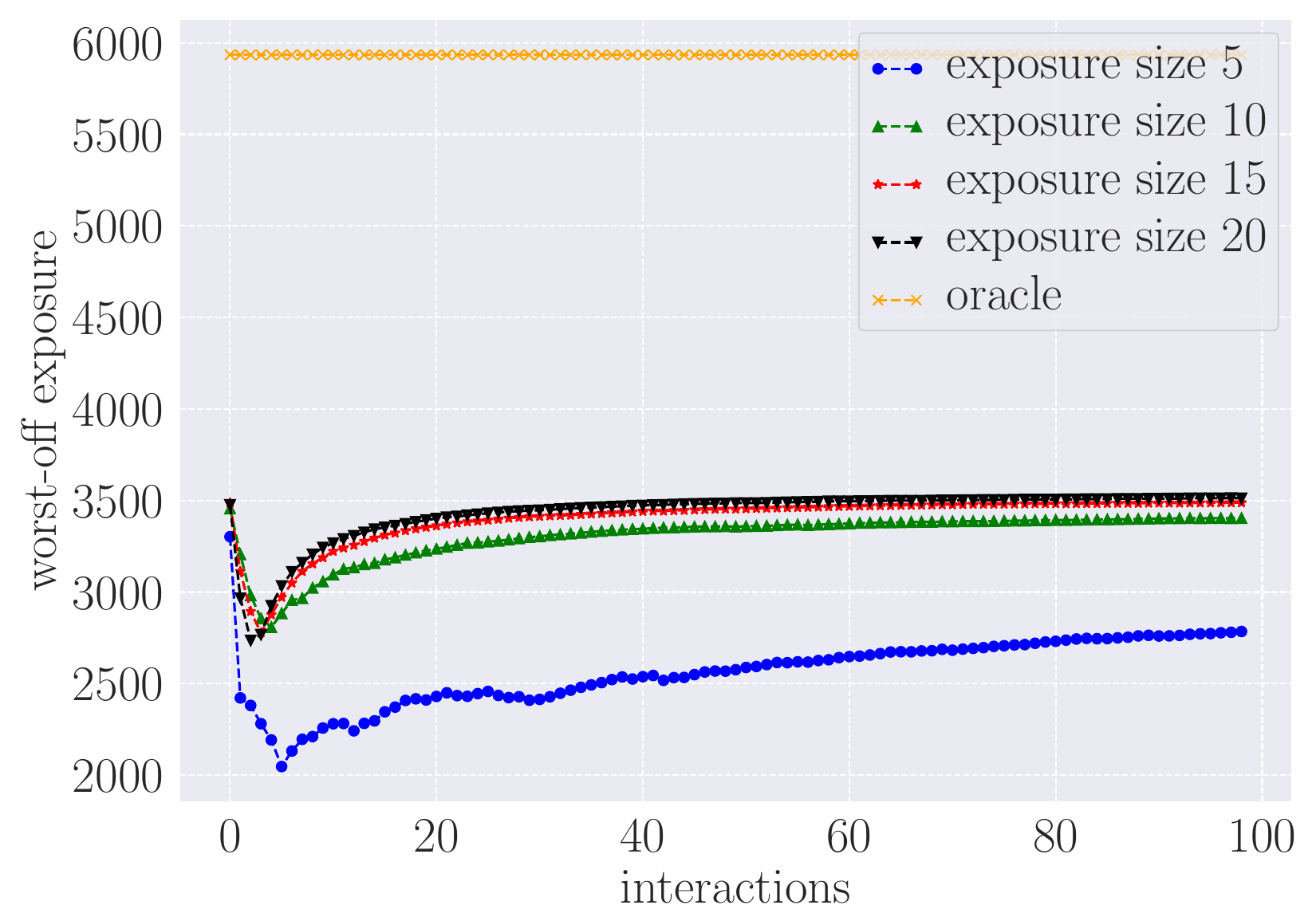}
        \label{fig:intro_simulation}
    }  
    \caption{\subref{fig:intro_loop} The feedback loops of the interaction between the fairness model and users.\subref{fig:intro_simulation}Simulations of the long-term lowest exposures among all provider (abbreviated as Lowest Exposures)}
    \label{fig:intro}
\end{figure}


In real scenarios, the components of recommender systems are not static. When conducting provider fairness, how to improve long-term performance under a time-changing environment is receiving increasing attention~\cite{mladenov2020optimizing, ge2021towards, ferraro2021break}. Previous work ~\cite{mladenov2020optimizing, ge2021towards, ferraro2021break} aimed to conduct long-term fairness under time-vary content providers, item popularity, and provider features, respectively. Different from these studies, we observed that recommendation feedback loops (RFL)~\cite{Survy:Unbias-Rec} will influence the provider max-min fairness greatly in the long term. As shown in Figure~\ref{fig:intro_loop}. The users and the RS are in RFL, where the model recommends a list of items to one user in each loop. Then, the user gives feedback, and the model will be updated incrementally according to the feedback. When updating the recommendation model, items without exposures are often viewed as negative samples (known as exposure/self-selection bias~\cite{Survy:Unbias-Rec}). Next, we will conduct an empirical study to describe how the RFL influences max-min fairness performance.

Figure~\ref{fig:intro_simulation} illustrates a simulation of how the unfairness increases in the repeated interactions between RS and users. We simulated $100$ interactions between $512$ users and an optimal fairness model, which can always obtain the optimal objective of the trade-off between predicting user preference and the max-min fairness objective. The preference estimation is updated using the users' feedback at the end of each interaction.
Every point in Figure~\ref{fig:intro_simulation} represents the long-term lowest exposures among all providers (abbreviated as Lowest Exposures) during the entire interaction process, and 
different lines represent the simulation performance under different exposure sizes (i.e. ranking size). The larger the exposure size becomes, the small providers will get more exposure opportunities~\cite{oosterhuis2020policy}. From Figure \ref{fig:intro_simulation}, one can easily observe that the magnitude of the exposure size (i.e. the impact of exposure bias) will significantly affect the long-term performance of the max-min fairness performance. The reason is that provider who has the lowest exposures will not get the opportunity to be exposed in such loop because his/her items will always be considered as negative samples. In RFL, providers with few exposures will get fewer and fewer opportunities to be exposed, intensifying the unfairness over time.


Although existing studies~\cite{swaminathan2017off,sun2019debiasing,jagerman2019people} propose to utilize reinforcement learning frameworks to alleviate the exposure bias amplification, how to optimize the long-term objective under RFL subject to fairness constraints is still a challenging problem. In this paper, we propose an online 
ranking model named Optimizing Long-Term Provider Max-min Fairness (LTP-MMF) to address the issue. Intuitively, LTP-MMF exploits the objective while exploring the feedback of unexposed items.

Specifically, LTP-MMF formulates the provider fair recommendation problem as a repeated resource allocation problem under batched bandit settings. Each item is considered a bandit's arm. At each round, when a user comes to access the RS, LTP-MMF will choose $k$ items (i.e. arms) to generate a recommendation ranking list. For each arm, 
LTP-MMF first utilizes the Matrix Factorization (MF) model~\cite{MF} to generate the predicted accuracy rewards from the side of users. Then, LTP-MMF will get provider exposures as the fairness reward in terms of the ranking list. Finally, LTP-MMF adds an exploration term toward the accuracy and fairness rewards by utilizing the Upper Confidence Bound (UCB) algorithm. Such exploration terms under fairness constraints help the model to access the feedback of unexposed items of small providers, improving the fairness performance in the long term.

As for the parameters of predicting rewards on the user's side, it has a serious problem of long update time. Therefore, we first collect enough users' feedback and then update them incrementally in a batched style. Theoretical
analysis shows that the regret of LTP-MMF enjoys a sub-linear bound and the batch size is a trade-off coefficient between accuracy and fairness in the long term.

We summarize the major contributions of this paper as follows:

(1)  In this paper, we analyze the importance of the long-term provider max-min fairness when considering the recommendation feedback loops;

(2) We formulate the long-term provider fair recommendation problem as a repeated resource allocation problem under a batched bandit setting and a ranking model called LTP-MMF is proposed. Theoretical analysis shows that the regret of LTP-MMF can be bounded;

(3) Extensive experiments on three public datasets demonstrated that LTP-MMF steadily outperforms the baselines in the long term. Moreover, we verified that LTP-MMF is computationally efficient in the online inference phase.

\section{Related Work}

The fairness problem in multi-stakeholder recommender systems has become a hot research topic~\cite{abdollahpouri2020multistakeholder,abdollahpouri2019multi}. According to different stakeholders, the fairness problem can be divided into customer fairness (C-fairness) and provider fairness (P-fairness)~\cite{burke2018balanced, cpfair}. In a recent publication, the stream of provider fairness is defined as ensuring that the item exposures of each provider are relatively similar to each other~\cite{qi2022profairrec, P-MMF, cpfair, wu2021tfrom, karimi2023provider, gomez2022provider, Dinnissen_FairInterview, fairrec, xu2024fairsync}. In generative recommender systems, provider fairness can also entail generating non-discriminatory item content for providers based on large language models~\cite{deldjoo2024understanding, li2023preliminary}. In this paper, our primary focus lies on mainstream provider fairness, which emphasizes maintaining as much equality as possible in the item exposures across different providers.

When dealing with provider fairness, most methods are conducted under re-ranking scenarios~\cite{P-MMF, cpfair, wu2021tfrom, karimi2023provider, gomez2022provider, xu2024fairsync, nips21welf, fairrec}. For example, FairRec~\cite{fairrec} and its extension FairRec+~\cite{fairrecplus} proposed an offline recommender model to guarantee equal frequency for all items in a series of ranking lists. Welf~\cite{nips21welf} proposed the Frank-Welf algorithm to solve the provider fair problem. Some work~\cite{cpfair, wu2021tfrom, karimi2023provider, gomez2022provider, ben2023learning} proposed a Linear Programming (LP)-based method to ensure the group fairness. P-MMF~\cite{P-MMF}, FairSync~\cite{xu2024fairsync} proposed an online mirror gradient descent to improve the worst-off provider's exposures in the dual space. In this paper, our main objective is to mitigate the influence of feedback loops in mirror gradient descent approaches~\cite{P-MMF, xu2024fairsync}, thereby achieving long-term provider fairness.

Another line of research proposed that the recommendation system is dynamic, the attributes, features, and parameters can change as time goes on~\cite{Survy:Unbias-Rec, ferraro2021break, ge2021towards, mladenov2020optimizing}. Previous studies aimed to 
improve the long-term performance of fairness. \citet{ge2021towards} focused on the time-vary recommendation attributes and formulated the long-term problem as a Constrained Markov Decision Process. ~\citet{mladenov2020optimizing} proposed that content providers may leave if they cannot obtain enough support and formulated the problem as a constrained matching problem. \citet{akpinar2022long} studied the 
provider population distributions that will change in the long term and proposed an intervention adjustment method. There are also some models~\cite{bistritz2020my, mandal2022socially} proposed to utilize reinforcement learning techniques to optimize long-term fairness. However, these long-term-based fairness methods did not consider the impact of recommendation feedback loop (i.e. exposure bias) in the long term.

In recent years, many debiasing approaches~\cite{yu2020influence, liu2020general} are proposed to break the feedback loop. For example, ~\cite{yu2020influence} proposed to utilize the influence function to remedy the bias from the loop, and \citet{liu2020general} proposed to utilize uniform data to debias from the loop. At the same time, there are also some reinforcement learning methods to break the loop. Exploitation and exploration in bandit settings were proposed to solve the problem. ~\citet{li2010contextual,hlinUCB, wang2017interactive} all utilized Upper Confidence Bound (UCB) to explore the unexposed items. Other models~\cite{zhao2018deep, wang2018reinforcement} also utilized policy gradient methods to break the feedback loop. However, they do not consider provider max-min fairness constraints in the feedback loop.

In the causal inference literature~\cite{Li23Causal, sheth2022causal, he2023addressing, Xu22dually}, some methods try to alleviate exposure bias in RS. For example, some work~\cite{Li23Causal, sheth2022causal} suggested utilizing user-social networks to mitigate exposure bias, as users who are closely connected are more likely to share the same items with each other. Some work~\cite{he2023addressing} proposed to utilize propensity score to mitigate exposure bias. DEPS~\cite{Xu22dually} considers the user-social network and item side propensity score together. However, these methods do not account for feedback loop scenarios and cannot be applied to fairness-aware RS.

\section{Formulation}\label{sec:formulation}
In this section, we first define the notations in multi-stakeholder recommender systems. Then, we give the formal definition of amortized max-min fairness accordingly. Finally, we formulate the interactions between users and the recommender model as a bandit problem.

\begin{figure}
    \centering
    \includegraphics[width=\linewidth]{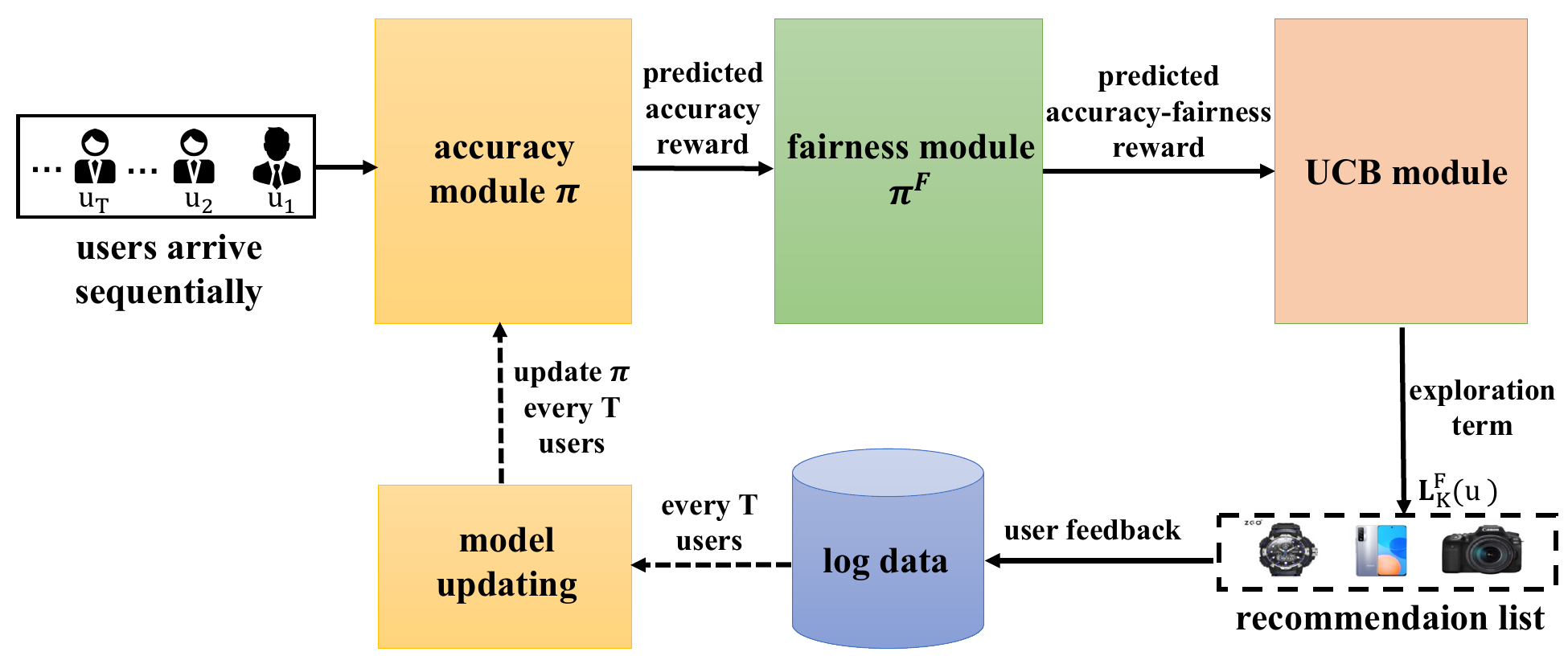}
    \caption{Sequential item ranking process of LTP-MMF}
    \label{fig:seq_procedure}
\end{figure}

\subsection{Multi-Stakeholders Recommender Systems}

A multi-stakeholder recommender system consists of different participants, including users, item providers, etc. Let $\mathcal{U}, \mathcal{I}$, and $\mathcal{P}$ be the set of users, items, and providers, respectively. Each item $i\in \mathcal{I}$ is associated with a unique provider $p\in \mathcal{P}$. The set of items associated with a specific provider $p$ is denoted by $\mathcal{I}_p$. 

In the ranking phase, when a specific user $u\in \mathcal{U}$ accesses the recommender system, for each user-item pair $(u, i)$, the click rate $s_{u,i}=P(c_{u,i}=1)$ of the user $u$ on the item $i$ is estimated through a ranking model based on the user's historical click feedback. We denote such feedback by $c_{u,i}\in\{0,1\}$, which represents whether or not the user clicked the item. The vector
$\bm{s}_u = [s_{u,1},s_{u,2},\cdots, s_{u,|\mathcal{I}|}]$ is the click rates of $u$ to all items. In the ranking phase, the items are ranked according to the estimated click rates. 


In provider-fair recommendation, we should consider the user-side utilities and provider-side fairness together. Formally, we define the true user-side utility of exposing item list $L_K(u)$ to $u$ as the summation of the preference scores in the list, denoted by $g\left(L_K(u)\right) = \sum_{i\in L_K(u)} s_{u,i}$, where $L_K(u)$ contains the $K$ items exposed to $u$. 
Following the literature convention \cite{fairrec,cpfair, P-MMF}, we define the fairness vector as $\mathbf{e}$, for a specific provider $p$, $\mathbf{e}_p\in\mathbb{R}^+$ denotes the exposure of provider $p$. In this paper, we mainly focus on provider max-min fairness~\cite{nips21welf,price4fairness, P-MMF}, which aims to improve the exposure opportunities of worst-off providers. Formally, $r(\bm{e}) = \min_{p\in|\mathcal{P}|}(\bm{e}/\gamma)$, where $\bm{\gamma}$ is the weighting vector of providers. 

In this paper, we aim to compute a fair list $L^F_K(u)\in\mathcal{I}^K$, which well balances the user utilities $g(\cdot)$ and provider fairness metric $r(\bm{e})$ under RFL in the long-term.




\subsection{Amortized Provider Max-Min Fairness}

In real-world applications, the users arrive at the recommender system sequentially. Assume that at time $t$ user $u_t$ arrives. The recommender system needs to consider long-term provider exposure during the entire time horizon from $t=0$ to $T$.
Our task can be formulated as a resource allocation problem with amortized fairness~\cite{balseiro2021regularized, P-MMF}. 
Specifically, the optimal utility of the recommender system can be defined as an amortized fairness function~\cite{balseiro2021regularized, P-MMF, biega2018equity}, which is the accumulated exposures over periods from 0 to $T$. In this case, $\mathbf{e}_p$ can be seen as the total number of exposed items of provider $p$, accumulated over the period $0$ to  $T$.

Formally, when trading off the user utilities and provider fairness,  we have the following mathematical program:
\begin{equation}
\label{eq:solve_OPT}
\begin{aligned}
         R_{OPT} = \max_{\mathbf{x}_t} \quad& \frac{1}{T}\sum_{t=1}^T g(\bm{x}_t) + \lambda r(\bm{e}) \\
         =\max_{\mathbf{x}_t} \quad& \frac{1}{T}\sum_{t=1}^T\sum_{i\in \mathcal{I}}\mathbf{x}_{ti}s_{u_t,i} + \lambda \min_{p\in|\mathcal{P}|}(\sum_{t=1}^T\bm{e}_t/\bm{\gamma})\\
        \textrm{s.t.}\quad  
        &\sum_{i\in\mathcal{I}}\mathbf{x}_{ti} = K, 
        \quad \forall t\in [1,2,\ldots,T]\\
         &\mathbf{e}_{t,p} = \sum_{i\in\mathcal{I}_p} x_{ti}, 
         \quad \forall p\in \mathcal{P}, \forall t\in [1,2,\ldots,T]\\
         &\sum_{t=1}^T\bm{e}_t \leq \bm{\gamma}, \quad \forall t\in [1,2,\ldots,T]\\
        & \mathbf{x}_{ti} \in \{0, 1\}, 
        \quad \forall i \in\mathcal{I}, ~\forall t\in [1,2,\ldots,T]
\end{aligned},
\end{equation}
where $\bm{\gamma}\in\mathbb{R}^{|\mathcal{P}|}$ denotes the weights of different providers, i.e., weighted MMF~\cite{price4fairness, P-MMF}. In amortized fairness, $\bm{e}\in\mathbb{R}^{T\times |\mathcal{P}|}$ is the exposure vector, i.e., $\bm{e}_{t,p}$ denotes the utility of provider $p$ at time $t$. $\mathbf{x}_t\in\{0, 1\}^{|\mathcal{I}|}$ is the decision vector for user $u_t$. Specifically, for each item $i$, $\mathbf{x}_{ti} = 1$ if it is in the fair ranking list $L_K^F(u_t)$, otherwise, $\mathbf{x}_{ti} = 0$. The max-min fairness regularizer $\min(\cdot)$ in the objective function suggests that we should improve the exposure opportunities of worst-off providers. The first constraint in Equation~\eqref{eq:solve_OPT} ensures that the recommended lists are of size $K$.
 The second constraint in Equation~\ref{eq:solve_OPT} suggests that the exposures of each provider $p$ are the accumulated exposures of the corresponding items over all periods. In general, we think time-separable fairness would be preferred under scenarios with weak timeliness. For example, recommending items with long service life (e.g., games and clothes, etc.).




Although here we have already given a linear programming solution Eq.\eqref{eq:solve_OPT} to the problem, it can only solve small-scale problems in an offline way. In online recommendation systems, for each user $u_t$ access, the model needs to generate a fair ranking list $L^F_K(u_t)$ from large-scale item corpus immediately. This means we have no idea about the information after $t$. Next, we will discuss how to use MMF in the online recommendation problem.

\subsection{Bandit with Provider Fairness}\label{sec:bandit}
We first define some notations for the problem. For any symmetric matrix $\bm{A}\in\mathbb{R}^{M\times M}$  and vector $\bm{x}\in\mathbb{R}^M$, let $\bm{x}_i$ denote the $i$-th element of the vector and $\bm{A}_i$ denote the $i$-th column of the matrix $\bm{A}$. Define $\|\bm{x}\|_{\bm{A}} = \sqrt{\bm{x}^{\top}\bm{A}\bm{x}}$. We also define the weighted $\ell_2$ norm to be $\|\bm{x}\|^2_{\bm{y}^2} = \sum_{i=1}^M \bm{x}_i^2\bm{y}_i^2$. 

To help the fairness model break the feedback loop, we formulate the provider-fair problem as a context bandit  process~\cite{wang2016learning}. As shown in
Figure~\ref{fig:seq_procedure}, at each iteration, a batch of users $\{u_t\}_{t=1}^T$ arrive sequentially. For each user $u_t$ and each item $i$, the accuracy module $\pi$ takes user features and item context as input and predicts the accuracy reward $\hat{s}_{u_t,i},\forall i\in\mathcal{I}$. Then the fairness module $\pi^F$ takes the accuracy reward as input and generates a predicted accuracy-fairness reward according to the provider exposures. Finally, the exploration module gives the exploration term according to previous recommendations to explore unexposed items. 

After getting the reward of each bandit arm (i.e., item), LTP-MMF chooses $K$ items with the highest rewards to generate an item list $L_K^F(u_t)$. Then the user gives their feedback $\{c_{u_t,i}, \forall i \in L_K^F(u_t)\}$, which is stored in the log. When a batch of users finishes their access, the recommendation results and rewards are collected.

In this paper, we propose to formulate such a process as a batched
bandit which can be represented by a 7-tuple $<\mathcal{S},\mathcal{L},\pi,\pi^F,R,N,T>$:

\textbf{Context space $\mathcal{S}$}:  denotes a hidden context space that summarizes the embedding space of both users and items. 

\textbf{Action space $\mathcal{L}$}:  denotes a given action space and each action corresponds to selecting $K$ items (arms): $L^F_K(u)\in\mathcal{L}$.

\textbf{Accuracy module  $\pi$}:  takes user and item context as input and predicts the accuracy reward if the user is recommended with this item, i.e., $\hat{s}_{u,i} = \pi(u,i)$.

\textbf{Fairness module $\pi^F$}: An online algorithm $\pi^F$ produces a real-time decision vector $\mathbf{x}_t\in \{0,1\}^{|\mathcal{I}|}$ based on the current user $u_t$ and the history
$\mathcal{H}_{t-1} = \{u_s,\mathbf{x}_s\}_{s=1}^{t-1}$. $\pi^F$ takes the estimated user-item preference score and item-provider relations as input and outputs an accuracy-fairness reward if the user is recommended with this item, i.e. $\hat{f}_{u,i} = \pi^F(u, \hat{s}_{u,i}, \mathcal{H}_{t-1})$.

\textbf{Reward $R$}:  is defined as a linear combination of the two types of feedback defined in Eq.~\eqref{eq:solve_OPT}: 
$
    R = \frac{1}{T}g(\bm{x}_t) + \lambda r(\bm{e}).
$

\textbf{Total data number $N$}: the process of a batched fairness-aware bandit is partitioned into $\lfloor N/T \rfloor$ episodes. Within each episode, the platform first updates the ranking policy $\pi$ using the collected user feedback and then applies the re-ranking policy $\pi^F$ with users for time horizon $T$ using the updated ranking policy $\pi$.

\textbf{Batch size $T$}: is the number of steps (time horizons) in each episode. That is, in each episode, the platform makes recommendations for each user and collects log data $\mathcal{B}=\{\bm{x}_t, c_{u_t,i},i\in L^F_K(u_t)\}, \bm{e}_t\}_{t=1}^T$. Finally, a
new accuracy predicted $\pi$ is trained on $\mathcal{B}$ at the beginning of the next episode.

Within each batch, the estimated reward is defined as 
$
\hat{R} = \frac{1}{T}\hat{g}(\bm{x}_t) + \lambda r(\bm{e})
$,
where $\hat{g}(\cdot),r(\cdot)$ are based on the estimated user-item score $\hat{s}$. The \emph{long-term regret} of the LTP-MMF is defined as the expectation over all log data $\mathcal{B}$:
\begin{equation}
    \text{Regret}\left(\pi,\pi^F\right) = \mathbb{E}_{\mathcal{B}}\left[R_{OPT}-\hat{R}\right].
\end{equation}

\section{Our approach: LTP-MMF}

In this section, we propose a novel recommender model which we call LTP-MMF. The model aims to optimize the long-term accuracy-fairness trade-off in recommendation loops. The whole procedure can be formulated as a batched bandit process in Figure~\ref{fig:seq_procedure}.

\subsection{Overall Architectures}
In this section, we will introduce the overall relations between the bandit settings discussed in Section~\ref{sec:bandit}. 
In the bandit problem, the overall reward $R$ will be divided into three parts. 
In the following three sections, we employ three modules—the Accuracy Module, the Fairness Module, and the UCB module—to compute three distinct parts of the reward, respectively.

\subsection{Accuracy Module: Matrix Factorization}\label{sec:MF}
For each $(u,i)$, we can define their hidden embeddings as $\bm{v}_u,\bm{v}_i \in \mathbb{R}^d$, where $d$ is the pre-defined dimension. The matrix factorization model takes two embeddings to determine the estimated preference score $\hat{s}_{u,i}$:
\begin{equation}
    \hat{s}_{u,i} = \pi(u,i) = \bm{v}_u^{\top}\bm{v}_i + \epsilon,
\end{equation}
where the random noise $\epsilon$ is drawn from a zero-mean Gaussian distribution $\mathcal{N}(0,\sigma^2)$.

The MF~\cite{koren2009MF} model uses a coordinate descent algorithm built upon the ridge regression to estimate the unknown parameter $\bm{v}_u$ for each user and the unknown hidden feature $\bm{v}_i$ for each item. Specifically, the objective function of the ridge regression can be written as follows
\begin{equation}\label{eq:MF}
    \min_{\bm{v}_u,\bm{v}_i} \frac{1}{2}\mathbb{E}_{i,u}\left[\left(\bm{v}_u^{\top}\bm{v}_i - c_{u,i}\right)^2 + \frac{\lambda_u}{2}\|\bm{v}_u\|_2 + \frac{\lambda_i}{2}\|\bm{v}_i\|_2\right],
\end{equation}
where where $\lambda_u$ and $\lambda_i$ are the trade-off parameters for the $\ell_2$ regularization.

In real-world applications, the users arrive at the recommender system sequentially. Assume that at time $t$ user $u_t$ arrives, the recommender model recommendes list $L_K(u_t)$ to $u_t$ based on estimated scores and receives the user click feedback $\{c_{u_t,i}, i\in L^t_K(u)\}$.
The closed-form estimation of $\bm{v}_i$, $\bm{v}_u$ at time $t$ can be derived as $\hat{\bm{v}}_{u,t} = (\bm{A}_{u,t})^{-1}\bm{b}_{u,t}$ and $\hat{\bm{v}}_{i,t}=(\bm{C}_{i,t})^{-1}\bm{d}_{i,t}$, where 
\begin{equation}\label{eq:MF_solve}
    \begin{aligned}
    \bm{A}_{u,t} = \lambda_i \bm{I} + \sum_{j=1}^t \hat{\bm{v}}_{i,j}(\hat{\bm{v}}_{i,j})^{\top}, \quad \bm{b}_{u,t} = \sum_{j=1}^t \hat{\bm{v}}_{i,j}c_{u,i};\\
    \bm{C}_{i,t} = \lambda_u \bm{I} + \sum_{j=1}^t \hat{\bm{v}}_{u,j}(\hat{\bm{v}}_{u,j})^{\top}, \quad \bm{d}_{i,t} = \sum_{j=1}^t \hat{\bm{v}}_{u,j}c_{u,i}.
\end{aligned}    
\end{equation}

\subsection{Fairness Module}
In this section, we will take the most state-of-the-art dual gradient descent method P-MMF~\cite{P-MMF} as our fairness module. It is important to note that our methods can be readily applied to various score-based re-ranking methods~\cite{xu2024fairsync, balseiro2021regularized, cpfair}, as our model LTP-MMF can easily substitute the fairness reward of the fairness module with other fairness rewards.
we aim to describe how to trade off user accuracy and provider fairness in an online fashion. Given the ranking scores from accuracy reward $s_{u,i}$, we can consider its dual problem:

\begin{theorem}[Dual Problem]\label{theo:dual}
The dual problem of Equation~\eqref{eq:solve_OPT} can be written as:
\begin{equation}
\label{eq:dual}
\begin{aligned}
        W_{OPT} &\leq W_{Dual} = \min_{\bm{\mu}\in\mathcal{D}}\left[g^*(\bm{M}\bm{\mu}) + \lambda r^*(-\bm{\mu})\right],
\end{aligned}
\end{equation}
where  $\bm{M}\in\mathbb{R}^{|\mathcal{I}|\times|\mathcal{P}|}$ is the item-provider adjacency matrix with $M_{ip} = 1$ indicating item $i\in \mathcal{I}_p$, $g^*(\cdot),r^*(\cdot)$ the conjugate functions:
\begin{equation}
    \begin{aligned}
    g^*(c) = \max_{\mathbf{x}_t\in\mathcal{X}}\sum_{t=1}^T\left[g(\mathbf{x}_t)/T - \mathbf{c}^\top\mathbf{x}_t\right], 
    r^*(-\boldsymbol{\mu}) = \max_{\mathbf{e}\leq \bm{\gamma}}\left[r(\mathbf{e})+\boldsymbol{\mu}^\top\mathbf{e}/\lambda\right]
    \end{aligned}
\end{equation}
with $\mathcal{X} = \{\mathbf{x}_t\mid\mathbf{x}_t \in \{0,1\} \land \sum_{i\in\mathcal{I}} \mathbf{x}_{ti} = K\}$,
and 
$\mathcal{D} = \{\boldsymbol{\mu}|r^*(-\boldsymbol{\mu})<\infty\}$ is the feasible region of dual variable $\boldsymbol{\mu}$. 

Moreover, the feasible region of the dual problem is
\begin{equation}
    \mathcal{D}_{\bm{\mu}} = \left\{\boldsymbol{\mu} ~~\left|~~ \sum_{p\in\mathcal{S}} \bm{\gamma}_p\boldsymbol{\mu}_p \ge -\lambda, \forall \mathcal{S}\in\mathcal{P}_s\right.\right\},
\end{equation}
where $\mathcal{P}_s$ is the power set of $\mathcal{P}$, i.e., the set of all the subsets of $\mathcal{P}$.
\end{theorem}

\begin{lemma}\label{lem:reg_form}
The conjugate function $r^*(\cdot)$ has a closed form:
\[
    \max_{\boldsymbol{\mu}\leq\boldsymbol{\gamma}}r^{*}(-\boldsymbol{\mu}) = \bm{\gamma}^T\boldsymbol{\mu}/\lambda + 1,
\]
and the optimal dual variable is:
\[
    \argmax_{\boldsymbol{\mu}\leq\boldsymbol{\gamma}}r^{*}(-\boldsymbol{\mu}) = \bm{\gamma}/\lambda.
\]
\end{lemma}

\begin{remark}
    Following the practice in~\cite{P-MMF}, we also try to optimize the integral linear programming from the dual space. It involves a vast variable $\bm{x}$'s space of size $T\times |\mathcal{I}|$ to dual variable $\mu$'s size \(|\mathcal{P}| \ll T \times |\mathcal{I}|\), and thanks to the sparsity of \(\mathbf{M}\), the computation of \(\mathbf{M}\bm{\mu}\) is highly efficient. Note that any other dual transformation way~\cite{xu2024fairsync, balseiro2021regularized} can be also applied in LTP-MMF.
\end{remark}

The proofs of Theorem~\ref{theo:dual} and Lemma~\ref{lem:reg_form} are provided in Appendix~\ref{app:dual_prove}. From Theorem~\ref{theo:dual}, we can have a new non-integral decision variable $\boldsymbol{\mu} \in \mathbb{R}^{|\mathcal{P}|}$. In practice, we usually have $|\mathcal{P}|\ll |\mathcal{I}|$. Besides, due to $\bm{M}$'s sparsity, it is very efficient to compute $\bm{M}\boldsymbol{\mu}$, which aims to project the variable $\boldsymbol{\mu}$ from the provider space onto the item spaces. According to Theorem~\ref{theo:dual}, the accuracy-fairness reward of a user-item pair is obtained through the conjunction function $g^*(\bm{M}\bm{\mu})$: $\hat{f}_{u,i} = \hat{s}_{u,i} - \bm{M}_i^{\top}\bm{\mu}$.

For each time step $t$ in a batch, we utilize the mirror momentum gradient descent~\cite{qian1999momentum, balseiro2021regularized} to learn the dual variable $\bm{\mu}_t$: first, we evaluate the conjugate function of the max-min regularizer $r^*(-\boldsymbol{\mu}_t)$ according to Lemma~\ref{lem:reg_form}. Then we can get the subgradient of the estimated dual function as
\[
    -\mathbf{\bm{M}}^{\top}\mathbf{x}_t + \mathbf{e}_t \in \partial \left(g^*(\bm{M}\bm{\mu}_t)+\lambda r^*(-\boldsymbol{\mu}_t)\right).
\]
Finally, we utilize $\mathbf{g}_t= \bm{M}\mathbf{x}_t + \bm{e}_t$ to update the dual variable by performing the online descent. Therefore, the dual variable will move towards the directions of the providers with fewer exposures, and the primal variable $\mathbf{x}_t$ will move to a better solution.


Note that the projection step can be efficiently solved using QP solvers~\cite{balseiro2021regularized} since $\mathcal{D}$ is coordinate-wisely symmetric. 
\begin{equation}
    \begin{aligned}
    \bm{\mu}_{t+1}  &= \argmin_{\bm{\mu}}{\sum_{p\in\mathcal{P}}}(\boldsymbol{\mu}_p\boldsymbol{\gamma}_p - \widetilde{\boldsymbol{\mu}}_{t,p}\boldsymbol{\gamma}_p)^2\\
    &\textrm{s.t.} \sum_{j=1}^m \boldsymbol{\gamma}_j\widetilde{\boldsymbol{\mu}}_j + \lambda \ge 0,~ \forall m = 1, 2 , \ldots, |\mathcal{P}|,  
    \end{aligned}    
\end{equation}

where $\widetilde{\boldsymbol{\mu}}$ satisfies:
\[
    \boldsymbol{\gamma}_1\widetilde{\boldsymbol{\mu}}_1 \leq \boldsymbol{\gamma}_2\widetilde{\boldsymbol{\mu}}_2 \leq \cdots, \leq \boldsymbol{\gamma}_{|\mathcal{P}|}\widetilde{\boldsymbol{\mu}}_{|\mathcal{P}|}.
\]

\begin{algorithm}[t]
    \caption{Online learning of LTP-MMF}
	\label{alg:bandit-MMF}
	\begin{algorithmic}[1]
	\REQUIRE User arriving order $\{u_i\}_{i=1}^N$, batch size $T$, ranking size $K$. 
    \textbf{Ranking parameters:} User embedding $\{\bm{v}_{u}\in\mathbb{R}^d,\forall u \in \mathcal{U}\}$, item embedding $\{\bm{v}_{i}\in\mathbb{R}^d,\forall i \in \mathcal{I}\}$. Confidence level $\sigma$, and semi-positive matrix $\bm{A}_u = \lambda_u\bm{I},\bm{b}_u = \bm{0}$, $\bm{C}_i = \lambda_i\bm{I},\bm{d}_i = \bm{0} $.
    
    \textbf{Re-ranking parameters:} Item-provider adjacent matrix $\mathbf{M}$, maximum resources $\bm{\gamma}$ and the coefficient $\lambda$. 
	\ENSURE The decision variables $\{\mathbf{x}_i, i = 1,2,\ldots, N\}$.
	\FOR{$n=1,\cdots,\lfloor N/T \rfloor$}
	    \STATE Initialize dual solution $\boldsymbol{\mu}_1 = 0$, remaining resources $\boldsymbol{\beta}_1 = \bm{\gamma}$, and momentum gradient $\mathbf{g}_0 = 0$, training buffer $\mathcal{B} =\emptyset$.
	    \FOR{$t=1,\cdots,T$}
    	    \STATE User $u_{nT+t}$ arrives, we abbreviate $u_{nT+t}$ as $u$.
                \STATE fairness model->users edge in Figure~\ref{fig:intro_loop}: lines 6-11.
        	\STATE Compute estimated score $s_{u,i} = \bm{v}_{u}^{\top}\bm{v}_i, \forall i\in\mathcal{I}$. 
                \STATE Compute the UCB $\triangle \bm{s}$ following Eq.~\eqref{eq:cr4rankning}:
                $
                    \triangle f_{u,i}^n = \alpha_{n} (\|v_i\|_{\bm{A}^{-1}_{u}}+C/2) + \beta_{n} (\|v_{u}\|_{\bm{C}^{-1}_i}+C/2).
                $
                \STATE
    	    $
            \mathbf{m}_p= (1-\bm{I}(\boldsymbol{\beta}_{tp} > 0) * 1000.0)$
            
            \STATE $// ~~ \text{Rewards for exploit-explore trade-off:}$
            \STATE
            $
            r_{u,i}= 
                \hat{s}_{u,i}/T-\mathbf{M}_i^{\top}(\boldsymbol{\mu}_{t}+\mathbf{m}) + \triangle f_{u,i}^n
            $
            \STATE yields output variable $\bm{x}$
    	    \STATE 
    	    $
    	        \mathbf{x}_{t} = \argmax_{\mathbf{x}_{t}\in\mathcal{X}}\left[\bm{r}_{u}^{\top}\bm{x}_t\right] 
    	    $
                \STATE users->data edge in Figure~\ref{fig:intro_loop}: lines 13-20.
                \STATE $\mathcal{B} = \mathcal{B}\cup \{(u,i,c_{u,i})$ $// ~~ \text{Receive user's feedback:}$
    	    \STATE$// ~~ \text{Update the remaining resources:}$
    	    \STATE $\boldsymbol{\beta}_{t+1} = \boldsymbol{\beta}_{t} - \bm{M}^{\top}\mathbf{x}_t, \mathbf{e}_t = \argmax_{\mathbf{e_t}\leq\boldsymbol{\beta}_t}r^*(-\mathbf{\boldsymbol{\mu}_t})$
            \STATE
            $
                \widetilde{\mathbf{g}}_t = -\mathbf{M}^{\top}\mathbf{x}_t + \mathbf{e}_t,
            $
            \STATE
            $ \mathbf{g}_t =\alpha \widetilde{\mathbf{g}}_t + (1-\alpha)\mathbf{g}_{t-1}
            $
            \STATE $// ~~ \text{Update the dual variable by gradient descent}$
            \STATE
            $
                \boldsymbol{\mu}_{t+1} = \argmin_{\boldsymbol{\mu}\in\mathcal{D}}\left[\langle \mathbf{g}_t,\boldsymbol{\mu} \rangle + \eta \| \boldsymbol{\mu}-\boldsymbol{\mu}_t \|_{\boldsymbol{\gamma}^2}^2\right]
            $
    	\ENDFOR
             \STATE data->fairness model edge in Figure~\ref{fig:intro_loop} lines 22-27.
            \FOR{$(u,i,c_{u,i})\in\mathcal{B}$}
            \STATE$// ~~ \text{Update accuracy module:}$
            \STATE 
            $\bm{A}_u = \bm{A}_u + \bm{v}_i\bm{v}_i^{\top},~\bm{C}_i = \bm{C}_i + \bm{v}_u\bm{v}_u^{\top}$
            \STATE
            $\bm{v}_u = \bm{A}_u^{-1}\bm{b}_u$, $\bm{v}_u = \bm{v}_u/\|\bm{v}_u\|_2$
            \STATE 
            $\bm{b}_u = \bm{b}_u + \bm{v}_ic_{u,i},~\bm{d}_i = \bm{d}_i + \bm{v}_uc_{u,i}$
            \STATE
            $
                \bm{v}_i = \bm{C}_i^{-1}\bm{d}_i, \bm{v}_i = \bm{v}_i/\|\bm{v}_i\|_2
            $
            \ENDFOR
	\ENDFOR
	\
	\end{algorithmic}
\end{algorithm}

\subsection{UCB for Accuracy-fairness Reward}

Upper Confidence Bound (UCB)~\cite{wang2017factorization,wang2016learning} has proven to be an effective strategy to estimate the
confidence of predicted rewards during exploration.
In fairness-aware recommendations, the system not only should exploit the accuracy-fairness objective in the ranking process but also should explore less exposed items to avoid giving too few exposures to some providers.


Next, we will bound the corresponding upper confidence bounds of the predicted accuracy-fairness reward $\hat{f}_{u,i}$. 
\begin{theorem}[Confidence Radius of Ranking]\label{theo:mf_ucb}
    The parameter $\bm{v}_u, \bm{v}_i$ is $q$-linear to the optimizer. For any $\sigma > 0$, with probability at least $1-\sigma$, the confidence radius $\triangle f_{u,i}^t$ of user-item preference score at time $t$ satisfies:
    \begin{equation}\label{eq:cr4rankning} 
    f_{u,i} - \hat{f}_{u_t,i} \leq \triangle f_{u,i}^t = \alpha_{t}(\|\hat{\bm{v}}_{i,t}\|_{\bm{A}_{u,t}^{-1}}+C_t/2) + \beta_{t}(\|\hat{\bm{v}}_{u,t}\|_{\bm{C}_{i,t}^{-1}}+C_t/2).
    \end{equation}
    We can decompose the confidence radius into three terms: 
    \begin{itemize}
        \item $\bm{v}_u$ and $\bm{v}_i$ bias terms $\|\hat{\bm{v}}_{u,t}-v_u^*\|_{\bm{A}_{u,t}},\|\hat{\bm{v}}_{i,t}-v_i^*\|_{\bm{C}_{i,t}}$ and their upper bound at time $t$ are denoted by $\alpha_{t},\beta_{i,t}$, respectively.
        \item $\bm{v}_u$ and $\bm{v}_i$ variance terms $\|\hat{\bm{v}}_{u,t}\|_{\bm{A}^{-1}_{u,t}},\|\hat{\bm{v}}_{i,t}\|_{\bm{C}^{-1}_{i,t}}$.
        \item collaborative variance terms $\|\hat{\bm{v}}_{u,t}-v_u^*\|_{\bm{A}^{-1}_{u,t}},\|\hat{\bm{v}}_{i,t}-v_i^*\|_{\bm{C}^{-1}_{i,t}}$ and their upper bound at time $t$ are defined as  $C_t = (q+\epsilon_q)^t$, where for any $\epsilon_q>0$ and $\bm{v}_u$, $\bm{v}_i$ are $q-$linear ($0<q<1$ to the optimizer.
    \end{itemize}
    The bias bound (i.e. exploration weight) $\alpha_u,\alpha_i$ are defined as 
    \begin{equation}\label{eq:bias_term}
     \begin{aligned}
        &\|\hat{\bm{v}}_{u,t}-{\bm{v}^*_{u}}\|_{\bm{A}_{u,t}} \leq 
 \alpha_{t},\quad \|\hat{\bm{v}}_{i,t}-{\bm{v}^*_{i}}\|_{\bm{C}_{i,t}} \leq \beta{t},\\ 
        &\alpha_{t} = \sqrt{\lambda_u} + \frac{2(q+\epsilon_q)(1-(q+\epsilon_q)^t)}{1-q-\epsilon_q} + \sqrt{d\ln{\frac{\lambda_ud+t}{\lambda_ud\sigma}}},\\
        &\beta_{t} = \sqrt{\lambda_i} + \frac{2(q+\epsilon_q)(1-(q+\epsilon_q)^t)}{1-q-\epsilon_q} + \sqrt{d\ln{\frac{\lambda_id+t}{\lambda_id\sigma}}}.
    \end{aligned}
    \end{equation}
The confidence radius can be bounded by
    $
    O\left(\frac{\left[1-(q+\epsilon_r)^t\right]\sqrt{\ln(t)}}{\sqrt{t}}\right),
    $
    which is a sub-linear decreasing function of t when $t$ becomes large. $O((1-(q+\epsilon_r)^t)\sqrt{\ln(t)})$ is the bias growth rate and $O(1/\sqrt{t})$ is the variance converge rate. 
\end{theorem}

\begin{remark}
    From Theorem 2, we can see the exploration term is $\triangle f_{u,i}^t = \alpha_{t}(\|\hat{\bm{v}}_{i,t}\|_{\bm{A}_{u,t}^{-1}}+C_t/2) + \beta_{t}(\|\hat{\bm{v}}_{u,t}\|_{\bm{C}_{i,t}^{-1}}+C_t/2)$. The exploration term has three parts: 
    
    $\bullet$ Bias terms $\|\hat{\bm{v}}_{u,t}-v_u^*\|_{\bm{A}_{u,t}},\|\hat{\bm{v}}_{i,t}-v_i^*\|_{\bm{C}_{i,t}}$ is bounded by $\alpha_t$, $\beta_t$, which captures the user preference shift with fairness. Intuitively, the presence of more shifts indicates that previous estimations may not be accurate, thus resulting in a larger degree of exploration.
    
    $\bullet$ Variance terms $\|\hat{\bm{v}}_{u,t}\|_{\bm{A}^{-1}_{u,t}},\|\hat{\bm{v}}_{i,t}\|_{\bm{C}^{-1}_{i,t}}$ captures the variance of user and item embedding estimation. Intuitively, the presence of more variance indicates that previous estimations may not be accurate, thus resulting in a larger degree of exploration.

    $\bullet$ Similarly, the collaborate variance term $C_t$ captures the variance of user-item interaction. Intuitively, the presence of more variance indicates that previous estimations may not be accurate, thus resulting in a larger degree of exploration.

    Those three parts together form the exploration term to balance exploration in fairness constraints with accuracy.
\end{remark}

The proof of theorem~\ref{theo:mf_ucb} is deferred to Appendix~\ref{app:mf_ucb_prove}. Equation~\ref{eq:cr4rankning} measures the estimation uncertainty of parameter $\bm{v}_u, \bm{v}_i$ in ranking module. The exploration and exploitation trade-off is balanced by the prediction confidence bound of both the user side and the item side. From Theorem~\ref{theo:mf_ucb}, we can also observe that the confidence radius is a trade-off between the bias term and the variance term. With more observations, the bias will increase but the variance will decrease more rapidly so that the radius bound
will become smaller (i.e. the bound is a sub-linear decreasing function of t when $t$ becomes large).

\subsection{Our Algorithm: LTP-MMF}

In each iteration of the algorithm, we first conduct online recommendations of user accuracy and provider max-min fairness. with fixed ranking model for a batch of users $\{u_t\}_{t=1}^T$. Then we will collect these users' feedback on the recommended items. Finally, the accuracy module parameters are updated using the collected users' feedback.

Algorithm~\ref{alg:bandit-MMF} illustrates our main algorithm.

In lines 6-11, the algorithm describes the process of fair-aware RS model giving recommendation results to users (fairness model->user edge in Figure 1). Specifically, when user $u$ comes to the recommender system, in line 6, LTP-MMF first estimates the accuracy score (accuracy Module). Then in line 7, the LTP-MMF computes the UCB bound (UCB module). in line 10, we will also add the final score with the fairness term (fairness module). 

These three modules together form the total reward: (1) For the error from the ranking model, we modify the ranking score according to its upper confidence bound. (2) For the max-min fairness objective, we apply the primal-dual theory to adjust the score with the dual variable $\bm{\mu}_t$. Intuitively, for $\bm{\mu}_t$, when the values of dual variables are higher, the algorithm naturally recommends fewer items related to the corresponding provider. (3) $\bm{m}$ ensures that the algorithm only recommends items from providers with remaining resources. Note that in line 9, the formulation is linear with respect to $\mathbf{x}_t$. Therefore, it is efficient to compute $\mathbf{x}_t$ through a top-$K$ sort algorithm in constant time. Finally, in line 12, we will adapt the top-K sorting operation to get the final output variable $\bm{x}$ according to the reward combined with these three components.

In line 14, the algorithm outlines the procedure by which users generate data for model training, specifically focusing on the user->data edge as depicted in Figure 1.

In lines 15-28, the algorithm outlines the process of model updating (data->fairness model edge in Figure 1). In lines 17-21, we aim to update the parameters of the fairness module utilizing the dual mirror descent algorithm. In lines 23-29, we utilize the user behavior data to update the parameters of the accuracy module.

\begin{theorem}[Regret Bound]\label{theo:regret}
Assume that the function $\|\cdot\|_{\boldsymbol{\gamma}^2}^2$ is $\sigma$-strong convex and there exists a constant $G,L\in\mathbb{R}^{+}$ such that $\|\widetilde{g}_t\|<G, \|\bm{\mu}\|_{\gamma}\leq L$.
Then, the regret can be bounded as follows:
\begin{equation}
    \text{Regret}(\pi,\pi^F) \leq \text{Regret}(\pi^F)+\text{Regret}(\pi)L,
\end{equation}
    where the regret bound of fairness module $\text{Regret}(\pi^F)$ is the regret of the dual fairness model, which can be bounded as
    \begin{equation}
    \text{Regret}(\pi^F)\leq \frac{N}{T}\left[\frac{K(1+\lambda \bar{r} + \bar{r})}{\min_p \boldsymbol{\gamma}_p} + \frac{L^2}{\eta} + \frac{G^2\eta(T-1)}{(1-\alpha)\sigma} + \frac{G^2}{2(1-\alpha)^2\sigma\eta}\right],
    \end{equation}
    where $\bar{r}$ is the upper bound of MMF regularzier, and in practice, $\bar{r}\leq 1$.
    The regret bound of accuracy module $\text{Regret}(\pi)$ can be bounded as 
    \begin{equation}
        \text{Regret}(\pi) \leq KT [\alpha_{\lfloor N/T \rfloor}(\rho_u+\kappa) + \beta_{\lfloor N/T \rfloor}(\rho_i+\kappa)],
    \end{equation}
    where 
    \[
    \kappa = \frac{(q+\epsilon_q)(1-(q+\epsilon_q)^{\lfloor N/T \rfloor})}{1-q-\epsilon_q},
    \]
    \[
    \rho_l = \sqrt{2d\frac{N}{T}\ln(1+\frac{N}{T\lambda_l d})}, l=u,i.
    \]
\end{theorem}

\begin{remark}[Accuracy-fairness regret trade-off]
    Setting the learning rate as $\eta  = O(T^{-1/2})$, we can obtain a fairness regret Regret($\pi^F$) upper bound of order  $O(N /\sqrt{T})$. The accuracy regret Regret($\pi$) is $O(\sqrt{NT\ln(\frac{N}{T})})$. The larger batch size $T$, the less error raised of accuracy in the long term, but the more bias caused by the fairness module $\pi^F$.
\end{remark}

\begin{remark}[Sublinear Long-term Regret]
    Overall, the long-term regret of LTP-MMF can be obtained from order $O(N\ln N)$. Moreover, we can observe that the more we tend to consider fairness, the larger $L$ will become, leading to larger regret in the long term.
\end{remark}

\subsection{Discussion}
We will summarize the key contribution of our method as follows. Firstly, our chosen accuracy module is the widely represented two-tower model of recommender systems~\cite{DMF}, which utilizes the dot product between two learned user item embeddings. However, it can be readily replaced with any other two-tower recommender system models~\cite{BPR, gru4recf, imporvedRec4gru} by simply substituting the learned embeddings for the user-item embeddings $v_i$ and $v_u$. Secondly, as mentioned in the Fairness Module, the fairness module can be also replaced by any score-based re-ranking methods~\cite{xu2024fairsync, balseiro2021regularized, cpfair}. Secondly, Our UCB exploration term exhibits wide adaptability, and our theoretical analysis also provides guidance for the application of other two-tower and score-based fairness models. Our method uniquely integrates the UCB technique to effectively balance accuracy estimation and fairness optimization within the feedback loop scenarios of RS, offering the literature a notable example of addressing the accuracy-fairness trade-off in such long-term contexts.

\section{Experiment}\label{sec:exp}
We conducted experiments to show the effectiveness of the proposed LTP-MMF in the long term.
The source code and experiments have been shared at github~\url{https://github.com/XuChen0427/LTP-MMF}.

\subsection{Experimental settings}
\subsubsection{Datasets}
The experiments were conducted on four large-scale, publicly available recommendation datasets, including:
 
 \textbf{Yelp}\footnote{\url{https://www.yelp.com/dataset}}: a large-scale businesses recommendation dataset. We only utilized the clicked data, which is simulated as the 4-5 star rating samples. 

 \textbf{Amazon-Beauty/Amazon-Baby}: Two subsets (beauty and digital music domains) of Amazon Product dataset\footnote{\url{http://jmcauley.ucsd.edu/data/amazon/}}. We only utilized the clicked data, which is simulated as the 4-5 star rating samples. Also, the brands are considered as providers. 
 
 \textbf{Steam}\footnote{\url{http://cseweb.ucsd.edu/~wckang/Steam_games.json.gz}}~\cite{SASRec}: We used the data for gamed played for more than 10 hours in our experiments. The publishers of games are considered as providers.

As a pre-processing step, the users, items, and providers who interacted with less than 5 items/users were removed from all datasets to avoid the extremely sparse cases. We removed providers associated with fewer than 5 items, which also included providers lacking a brand name, resulting in the removal of their associated items as well. We will give a more detailed provider-removing ratio in Table~\ref{tab:dataset}.
Table~\ref{tab:dataset} lists some statistics of the four datasets. 

\begin{table}[t]
\caption{Statistics of the datasets. } \label{tab:dataset}
\centering
\begin{tabular}{lrrrrrr}
\hline
Dataset              & \#User & \#Item & \#Provider &  \#Interaction & Sparsity & Provider Removing Ratio \\
\hline
\hline
Yelp                 & 17034  & 11821  & 23 & 154543      & 99.92\% & 0.0\%\\
Amazon-Beauty         & 9625   & 2756   & 104  & 49217    & 99.81\% & 50.55\%\\
Amazon-Baby         & 11680   & 2687   & 112  & 59836    & 99.80\% & 60.94\%\\
Steam         & 5902   & 591   & 81  & 29530    & 99.15\% & 25.06\%\\
\hline
\end{tabular}
\end{table}

\subsubsection{Evaluation}
We sorted all the interactions according to the time and used the first 80\% of the interactions as the training data to train the base model (i.e., BPR~\cite{BPR}). The remaining 20\% of interactions were used as the test data for evaluation. Based on the trained base model, we can obtain a preference score $s_{u, i}$ for each user-item pair $(u, i)$.
The chronological interactions in the test data were split into interaction sequences where the horizon length was set to $T$. We calculated the metrics separately for each sequence, and the averaged results are reported as the final performances.

As for the evaluation metrics, the performances of the models were evaluated from three aspects: user-side preference, provider-side fairness, and the trade-off between them. As for the user-side accuracy, following the practices in~\cite{wu2021tfrom}, we utilized the CTR@K, which measures the averaged click rate in the long term:
\begin{equation}
\text{CTR@K} =\frac{1}{N} \sum_{t=1}^N\sum_{i\in\mathbf{L}^F_K(u_t)}s_{u_t,i}/K.
\end{equation}

As for provider fairness, we directly utilized the definition of MMF in Section~\ref{sec:formulation} as the metric:
\begin{equation}
    \textrm{MMF}@K = \frac{T}{N}\sum_{i=1}^{\lfloor N/T \rfloor}\min_{p\in\mathcal{P}}\left\{\sum_{t=1}^T\sum_{i\in\mathbf{L}_K^F(u_t)}\mathbb{I}(i\in\mathcal{I}_p)/\boldsymbol{\gamma}_p\right\},
\end{equation}
where $\mathbb{I}(\cdot)$ is the indicator function. 

As for the trade-off performance, we used the online objective traded-off to measure the fairness:
\begin{equation}
    r_{\lambda}@K = \text{CTR@K} + \lambda\cdot \textrm{MMF}@K,
\end{equation}
where $\lambda\geq 0$ is the trade-off coefficient.

\subsubsection{Baselines}
The following representative provider fair re-ranking models were chosen as the baselines:
\textbf{FairRec}~\cite{fairrec} and \textbf{FairRec+} \cite{fairrecplus} aimed to guarantee at least Maximin Share (MMS) of the provider exposures.
\textbf{CPFair}~\cite{cpfair} formulated the trade-off problem as a knapsack problem and proposed a greedy solution.

We also chose the following MMF models:
\textbf{Welf}~\cite{nips21welf} use the Frank-Wolfe algorithm to maximize the Welfare functions of worst-off items. However, it is developed under off-line settings; \textbf{RAOP}~\cite{balseiro2021regularized} is a state-of-the-art online resource allocation method. We applied it to the recommendation by regarding the items as the resources and users as the demanders. \textbf{Fairco}~\cite{Fairco}: added a regularizer that measures the exposure gaps between the target provider and the worst providers.

We also compared the proposed with one heuristic MMF baseline:
\textbf{$K$-neighbor}: at each time step $t$, only the items associated to the top-$K$ providers with the least cumulative exposure are recommended.
At the same time, we also compared three bandit baselines, which aim to remedy the bias from the feedback loops of RS. Note that, to make a fair comparison, all the bandit algorithms are updated based on the batched log.
\textbf{hLinUCB}~\cite{wang2016learning}: it utilized UCB for learning hidden features for both users and items.
\textbf{EXP3}~\cite{EXP3} aimed to choose actions according to a distribution constructed by the exponential weights.
\textbf{BLTS}~\cite{BLTS} will sample a reward based on a normal distribution with the estimated reward variance.
However, these bandit baselines did not consider provider fairness in the long term.

Meanwhile, we also compare the baseline Dyn~\cite{akpinar2022long} of optimizing long-term fairness in feedback loops. However, Dyn addresses long-term fairness concerns by considering the dynamic nature of user social networks, where changes in social connections may lead users to belong to different social groups over time. Therefore, we apply their re-ranking methods to our framework by treating specific user groups as providers and users within those groups as items belonging to the respective provider. To ensure a fair comparison, we substitute their dynamic environment, which involved changing social networks, with dynamic user-item score estimation. 

\begin{table*}[ht]
\setlength{\tabcolsep}{7.5pt}
        \Small
        \caption{List of critical parameters related to fairness in Table~\ref{tab:EXP:main}. To ensure a fair comparison, if the baseline includes a linear trade-off coefficient, we uniformly set $\lambda=0.5$. For other parameters, we tuned them within the following ranges for each dataset.}
    \label{tab:EXP:paras}
    \centering
    \centering
\begin{tabular}{|cc|}
\toprule
models & fairness-aware  parameters \\
\hline
k-neighbor & $k\in[1,5]$ \\
P-MMF & $\lambda=0.5$ \\
CPFair & $\lambda=0.5$ \\
Fairco & $\lambda=0.5$ \\
FairRec/FairRec+ & $\alpha\in[0.5,1]$ \\
Welf & $\lambda=0.5$ \\
Dyn & $\lambda=0.5$  \\
\textbf{LTP-MMF(ours)} & $\lambda=0.5, q=0.8, \lambda_u=\lambda_i=1$ \\
\bottomrule
\end{tabular}
\end{table*}

\begin{table*}[ht]
\setlength{\tabcolsep}{5pt}
        \SMALL
        \caption{Performance comparisons between LTP-MMF and the baselines. The experimental settings are listed in Table~\ref{tab:EXP:paras}. The bold numbers denote the performance of our models, and the underlined numbers denote the best-performing baselines.} $*$: improvements over the best baseline are statistically significant ($t$-test, $p$-value< 0.05).
    \label{tab:EXP:main}
    \centering
    \centering
    \begin{tabular}{lrrrrrr|rrr}
       \toprule
       \multirow{2}{*}{Model} & \multicolumn{3}{c}{User Accuracy(CTR@K)} & \multicolumn{3}{c}
        {Provider Fairness(MMF@K)} 
        & \multicolumn{3}{c}{\textbf{Trade-off Performance(r@K)$^*$}} \\

\cmidrule(lr){2-4}\cmidrule(lr){5-7}\cmidrule(lr){8-10}

        & $CTR@$5 & $CTR@$10 & $CTR@$20  & $MMF@$5 & $MMF@$10 & $MMF@$20 
                & $r@$5 & $r@$10 & $r@$20 
        \\
        \hline
\multicolumn{10}{c}{\textbf{Yelp}}
\\
        \hline

hLinUCB  & 0.685 & 0.654 & 0.661 & 0.000 & 0.000  & 0.000 & 0.685 & 0.654 & 0.661\\

EXP3  & 0.463 & 0.465 & 0.466 & 0.014 & 0.022  & 0.028 & 0.470 & 0.466 & 0.480\\

BLTS  & 0.703 & 0.751 & \underline{0.770} & 0.002 & 0.005  & 0.005 & 0.703 & 0.753 & 0.773\\

\hline
FairRec  & 0.710 & 0.687 & 0.762 & 0.000 & 0.000  & 0.000 & 0.710 & 0.687 & 0.762\\

FairRec+ & \underline{0.717} & \underline{0.802} & 0.762 & 0.000 & 0.000  & 0.000 & \underline{0.717} & \underline{0.802} & 0.762\\

CPFair& 0.625 & 0.659 & 0.709 & 0.071 & 0.070 & 0.109 & 0.661 & 0.694 & 0.764\\

k-neighbor& 0.441 & 0.444 & 0.465 & 0.145 & 0.145  & 0.143 & 0.514 & 0.517 & 0.537\\

Welf & 0.556 & 0.577 & 0.607 & 0.158 & 0.160 & 0.182 & 0.635 & 0.657 & 0.698\\

Fairco& 0.569 & 0.587 & 0.606 & \underline{0.271} & \underline{0.278} & 0.283 & 0.705 & 0.726 & 0.748\\

P-MMF & 0.581 & 0.611 & 0.648 & 0.225 & 0.314  & \underline{0.390} & 0.694 & 0.768 & \underline{0.843}\\

Dyn & 0.507 & 0.504 & 0.464 & 0.252 & 0.134 & 0.001 & 0.663 & 0.571 & 0.4645 \\
\hline

\textbf{LTP-MMF (Ours)}& \textbf{0.621} & \textbf{0.665} & \textbf{0.704} & \textbf{0.386} & \textbf{0.363} & \textbf{0.327} & \textbf{0.814}$^*$ & \textbf{0.847}$^*$ & \textbf{0.868}$^*$\\

\hline
Improv. & &  &  & &  &  & 13.5\% & 5.6\% & 3.0\% \\

\hline
\hline

\multicolumn{10}{c}{\textbf{Amazon-Beauty}}
\\
        \hline

hLinUCB  & \underline{0.588} & \underline{0.601} & 0.565 & 0.000 & 0.000  & 0.000 & 0.588 & 0.601 & 0.565\\

EXP3  & 0.455 & 0.455 & 0.456 & 0.004 & 0.076  & 0.135 & 0.457 & 0.493 & 0.524\\

BLTS  & 0.559 & 0.566 & \underline{0.580} & 0.000 & 0.010  & 0.026 & 0.559 & 0.571 & 0.593\\

\hline
FairRec  & 0.556 & 0.575 & 0.557 & 0.000 & 0.000  & 0.110 & 0.556 & 0.575 & 0.612\\

FairRec+ & 0.556 & 0.575 & 0.559 & 0.000 & 0.000  & 0.111 & 0.556 & 0.575 & 0.615\\

CPFair& 0.541 & 0.549 & 0.558 & 0.006 & 0.013  & 0.095 & 0.544 & 0.556 & 0.606\\

k-neighbor& 0.453 & 0.469 & 0.499 & 0.125 & 0.133  & 0.156 & 0.516 & 0.536 & 0.577\\

Welf & 0.536 & 0.554 & 0.556 & 0.000 & 0.000 & 0.167 & 0.536 & 0.554 & 0.640\\


Fairco& 0.502 & 0.507 & 0.518 & 0.227 & 0.218  &0.229 & 0.616 & 0.616 & 0.633\\

P-MMF & 0.505 & 0.526 & 0.543 & \underline{0.452} & \underline{0.535}  & \underline{0.583} & \underline{0.731} & \underline{0.794} & \underline{0.835}\\

Dyn & 0.439 & 0.442 & 0.444 & 0.003 & 0.002 & 0.001 & 0.4405 & 0.443 & 0.4445 \\

\hline

\textbf{LTP-MMF (Ours)}& \textbf{0.505} & \textbf{0.531} & \textbf{0.553} & \textbf{0.468} & \textbf{0.545}  & \textbf{0.572} & \textbf{0.739}$^*$ & \textbf{0.804}$^*$ & \textbf{0.840}$^*$\\

\hline

Improv. &  & & & &  &  & 1.1\% & 1.3\% & 0.6\% \\

       \hline
       \hline

\multicolumn{10}{c}{\textbf{Amazon-Baby}}
\\
\hline

hLinUCB  & \underline{0.570} & 0.511 & 0.514 & 0.000 & 0.000  & 0.000 & 0.570 & 0.511 & 0.514\\

EXP3  & 0.463 & 0.463 & 0.463 & 0.040 & 0.094  & 0.138 & 0.483 & 0.510 & 0.532\\

BLTS  & 0.546 & \underline{0.549} & \underline{0.551} & 0.006 & 0.024  & 0.055 & 0.549 & 0.561 & 0.578\\

\hline
FairRec  & 0.519 & 0.531 & 0.546 & 0.000 & 0.000 & 0.087 & 0.519 & 0.531 & 0.590\\

FairRec+ & 0.519 & 0.531 & 0.549 & 0.000 & 0.000 & 0.086 & 0.519 & 0.531 & 0.592\\

CPFair& 0.524 & 0.536 & 0.536 & 0.017 & 0.046  & 0.104 & 0.533 & 0.559 & 0.588\\

k-neighbor& 0.464 & 0.472 & 0.487 & 0.144 & 0.150  & 0.173 & 0.536 & 0.547 & 0.574\\

Welf & 0.514 & 0.525 & 0.529 & 0.199 & 0.271  & 0.246 & 0.614 & 0.661 & 0.652\\

Fairco& 0.496 & 0.504 & 0.507 & 0.253 & 0.252  & 0.250 & 0.623 & 0.630 & 0.632\\

P-MMF & 0.493 & 0.512 & 0.521 & \underline{0.533} & \underline{0.560}  & \underline{0.580} & \underline{0.760} & \underline{0.792} & \underline{0.811}\\

Dyn & 0.455 & 0.452 & 0.451 & 0.032 & 0.016 & 0.006 & 0.471 & 0.46 & 0.454\\

\hline

\textbf{LTP-MMF (Ours)}& \textbf{0.498} & \textbf{0.515} & \textbf{0.526} & \textbf{0.534}$^*$ & \textbf{0.571}$^*$  & \textbf{0.595}$^*$ & \textbf{0.765}$^*$ & \textbf{0.801}$^*$ & \textbf{0.824}$^*$\\

\hline

Improv. &  &  &  & &  &  & 0.7\% & 1.1\% & 1.6\%        \\
        \hline
\hline

\multicolumn{10}{c}{\textbf{Steam}}
\\
        \hline

hLinUCB  & 0.820 & 0.574 & 0.707 & 0.000 & 0.000  & 0.000 & 0.820 & 0.574 & 0.707\\

EXP3  & 0.313 & 0.313 & 0.315 & 0.158 & 0.205  & 0.224 & 0.392 & 0.415 & 0.427\\

BLTS  & \underline{0.829} & \underline{0.849} & 0.018 & 0.011 & 0.017  & 0.835 & 0.834 & 0.857 & 0.844\\

\hline
FairRec  & 0.573  & 0.603 & 0.612 & 0.141 & 0.140  & 0.138 & 0.644 & 0.673 & 0.681\\

FairRec+ & 0.583 & 0.604 & 0.614 & 0.153 & 0.148  & 0.139 & 0.660 & 0.678 & 0.684\\

CPFair& 0.713 & 0.734 & \underline{0.756} & 0.013  & 0.016  & 0.046 & 0.720 & 0.742 & 0.779\\

k-neighbor&0.364 &0.408   & 0.597 & 0.127 &  0.179 & 0.220 & 0.428 & 0.498 & 0.707\\

Welf & 0.680 & 0.672 & 0.702 & 0.000 &  0.162 & 0.177 & 0.680 & 0.753 & 0.791\\


Fairco& 0.538 & 0.534 & 0.547 & 0.264 & 0.265  & 0.254 & 0.670 & 0.676 & 0.674\\

P-MMF & 0.573 & 0.593 & 0.624 & \underline{0.547} & \underline{0.602}  & \underline{0.608} & \underline{0.847} &  \underline{0.894} & \underline{0.928}\\
\hline

Dyn & 0.442 & 0.258 & 0.198 & 0.159 & 0.039 & 0.015 & 0.521 & 0.277 & 0.211\\

\textbf{LTP-MMF(Ours)}& \textbf{0.602}& \textbf{0.611} & \textbf{0.633} & \textbf{0.603} & \textbf{0.617}  & \textbf{0.628} & \textbf{0.904}$^*$ & \textbf{0.920}$^*$ & \textbf{0.947}$^*$\\

\hline

Improv. &  &  & &  &  &  & 6.7\% & 2.9\% & 2.0\%\\

\bottomrule

    \end{tabular}
\end{table*}

\clearpage

\subsubsection{Implementation details}\label{sec:Implementation}

As for the hyper-parameters in all models, the learning rate was tuned among $[1e-2,1e-3]/T^{1/2}$, and the momentum coefficient $\alpha$ was tuned among $[0.2,0.5]$. For the maximum resources (i.e., the weights) $\bm{\gamma}$, following the practices in~\cite{wu2021tfrom, P-MMF}, we set $\bm{\gamma}$ based on the number of items provided by the providers:
\begin{equation}
     \bm{\gamma}_p = KT\eta{|\mathcal{I}_p|}\big/{|\mathcal{I}|},
\end{equation}
where $\eta$ is the factor controlling the richness of resources. In all the experiments, we set $\eta=1+1/|\mathcal{P}|$.
We implemented LTP-MMF with PyTorch~\cite{pytorch}. The experiments were conducted with a single NVIDIA GeForce RTX 3090.

\begin{figure*}
    \centering
    \subfigure[Yelp]{
        \centering
        \includegraphics[width=0.48\linewidth]{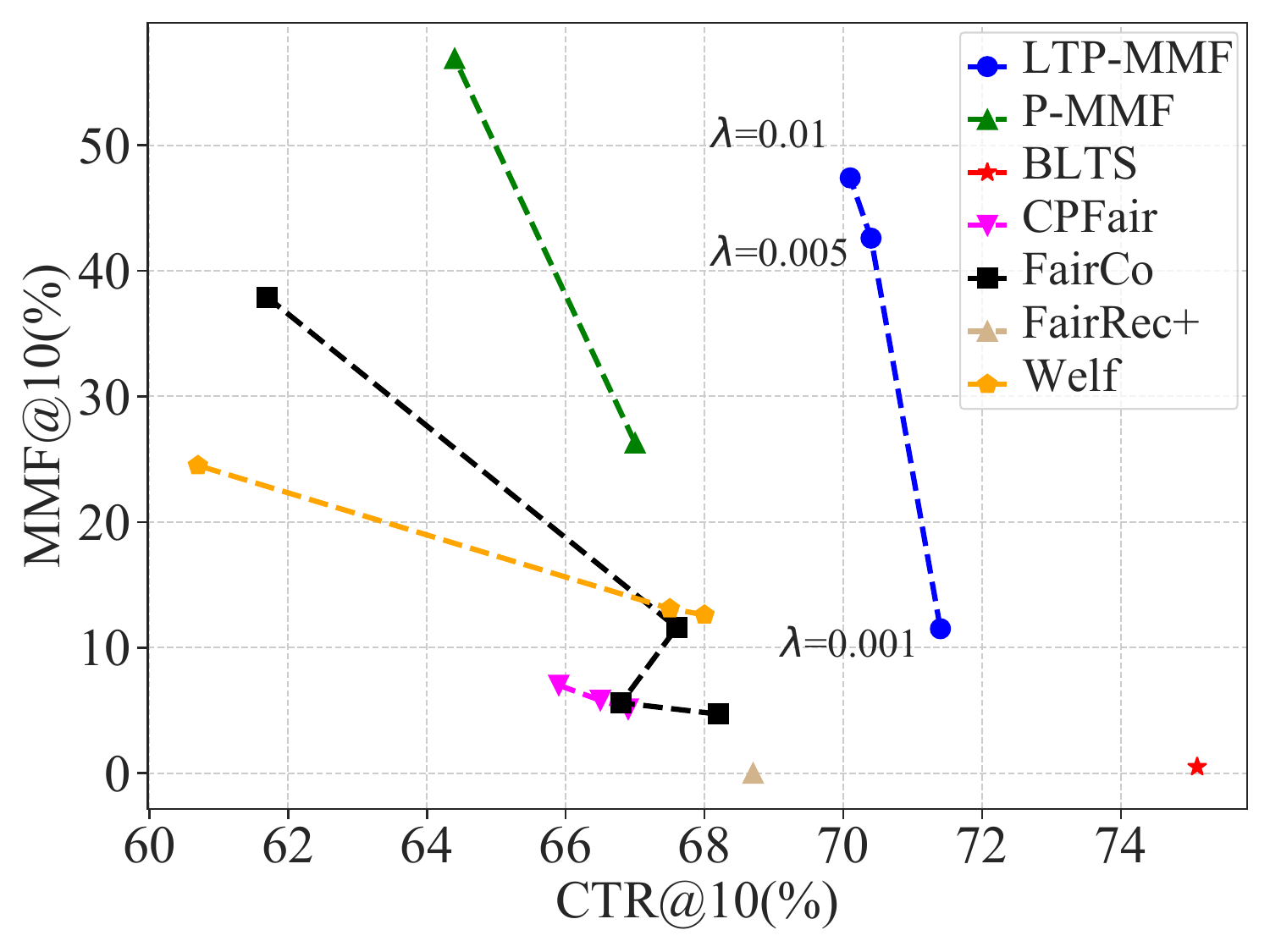}
    
        \label{fig:pareto_yelp}        
}
    \subfigure[Amazon Beauty]{
        \centering
        \includegraphics[width=0.48\linewidth]{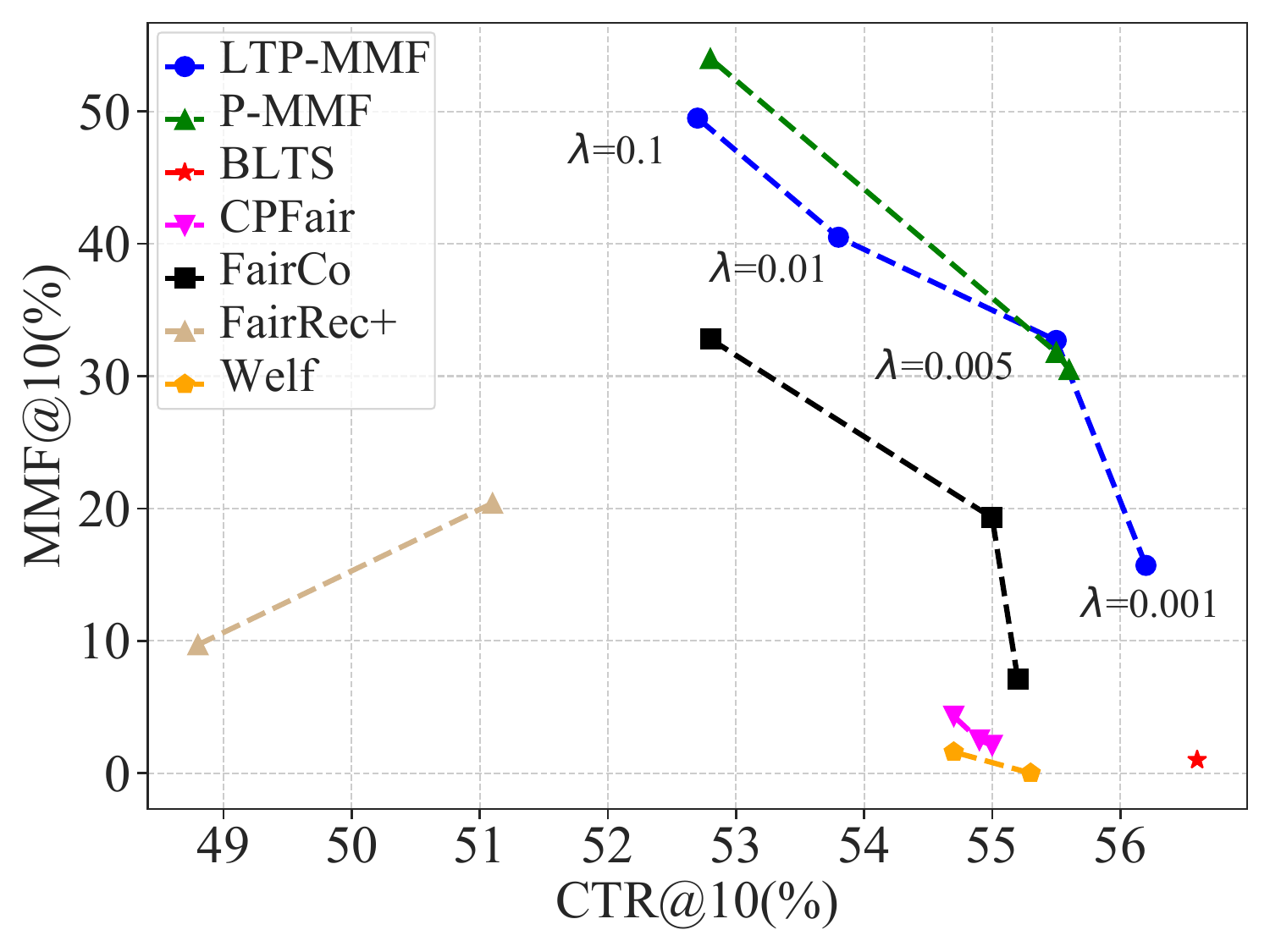}
        
        \label{fig:pareto_amazon_beauty}        
}
\subfigure[Amazon Baby]{
        \centering
        \includegraphics[width=0.48\linewidth]{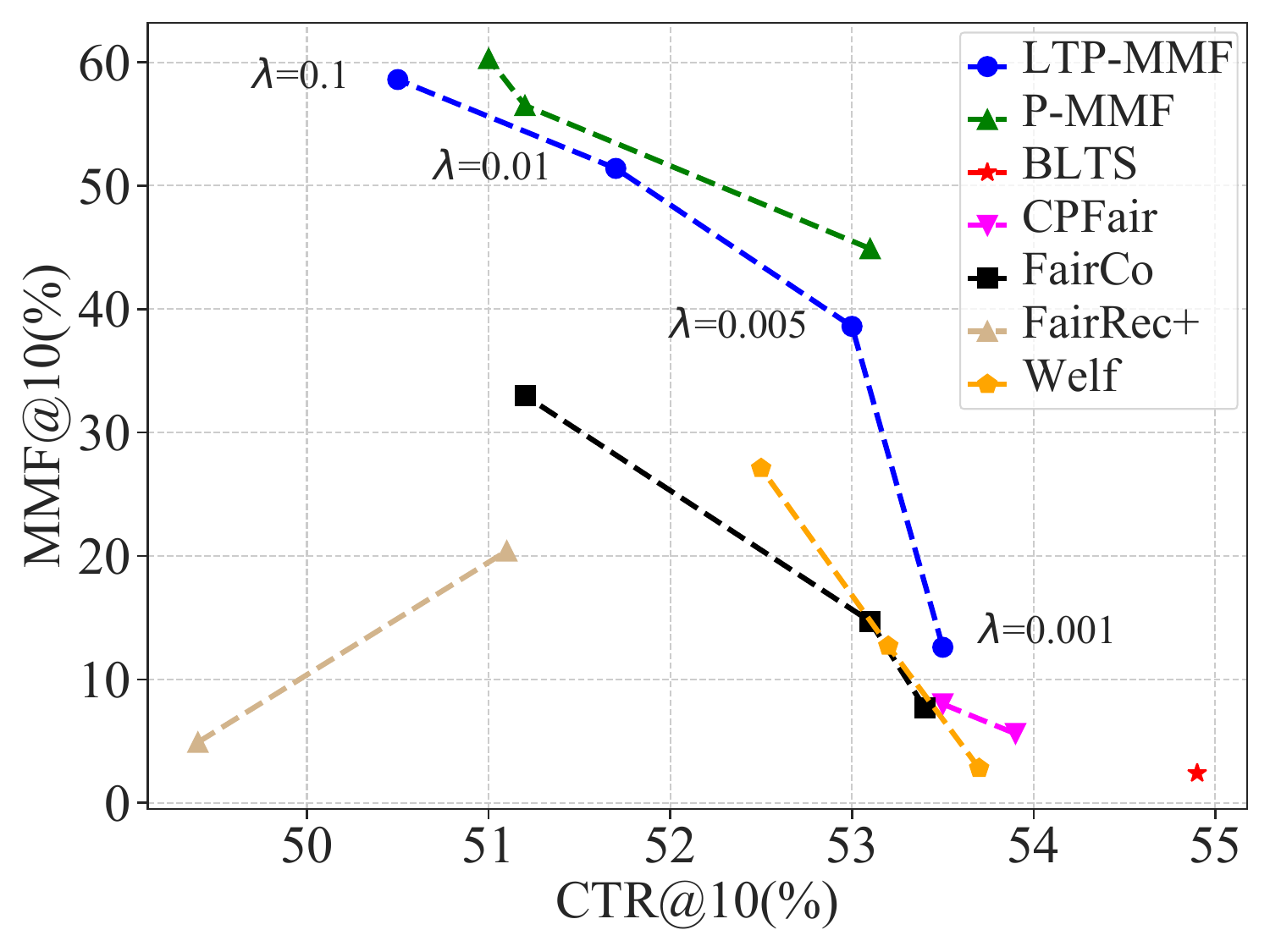}

        \label{fig:pareto_amazon_baby}        
}
    \subfigure[Steam]{
        \centering
        \includegraphics[width=0.48\linewidth]{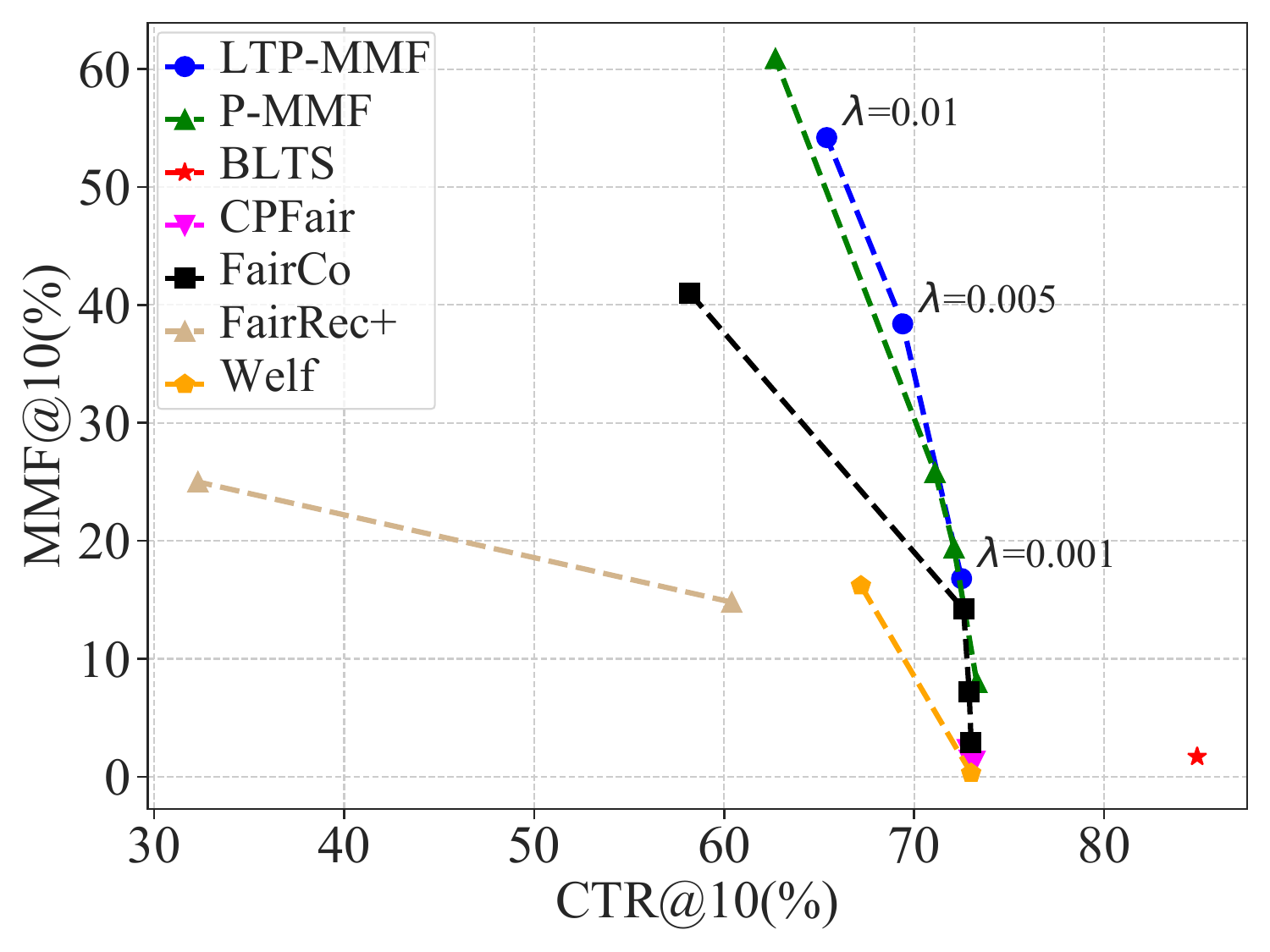}
        \label{fig:pareto_steam}        
}

    \caption{ Pareto frontier in Four Dataset
 }
 \label{fig:pareto}
\end{figure*}

\begin{figure}
        \centering
\includegraphics[width=0.9\linewidth]{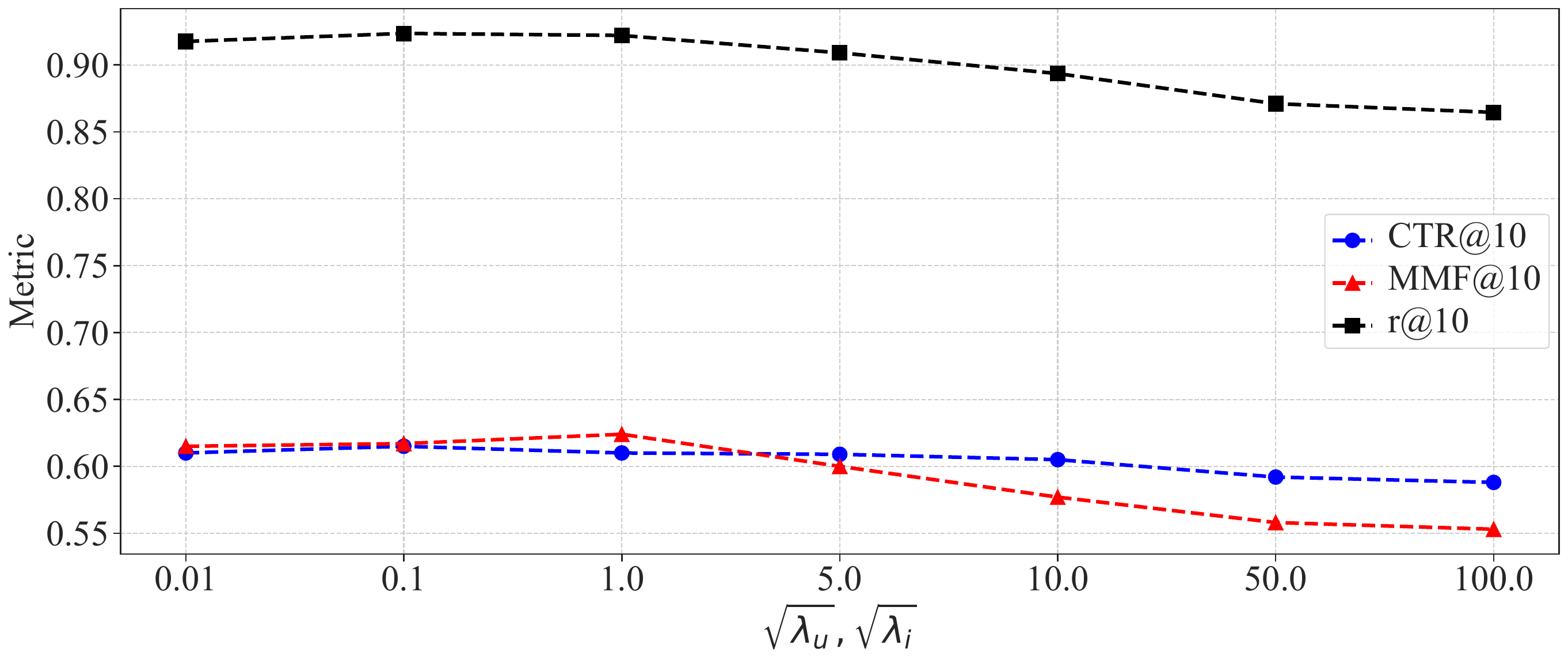}
        \caption{Exploration Weight}\label{fig:alpha_u} 
\end{figure}

\subsection{Main Experiments}
In this section, we conducted the experiments on four large-scale datasets with the LTP-MMF and other baselines. 

Due to variations in dataset characteristics, we will provide a comprehensive list of critical parameters related to fairness performance ranges for our model LTP-MMF and other baselines in Table~\ref{tab:EXP:paras}. To make a fair comparison, the other important variables such as $\gamma$ for evaluation and the learning rate are all listed in Section~\ref{sec:Implementation} in the revised version.




\begin{figure}
    \centering
    \subfigure[CTR]
{
        \centering
        \includegraphics[width=0.48 \linewidth]{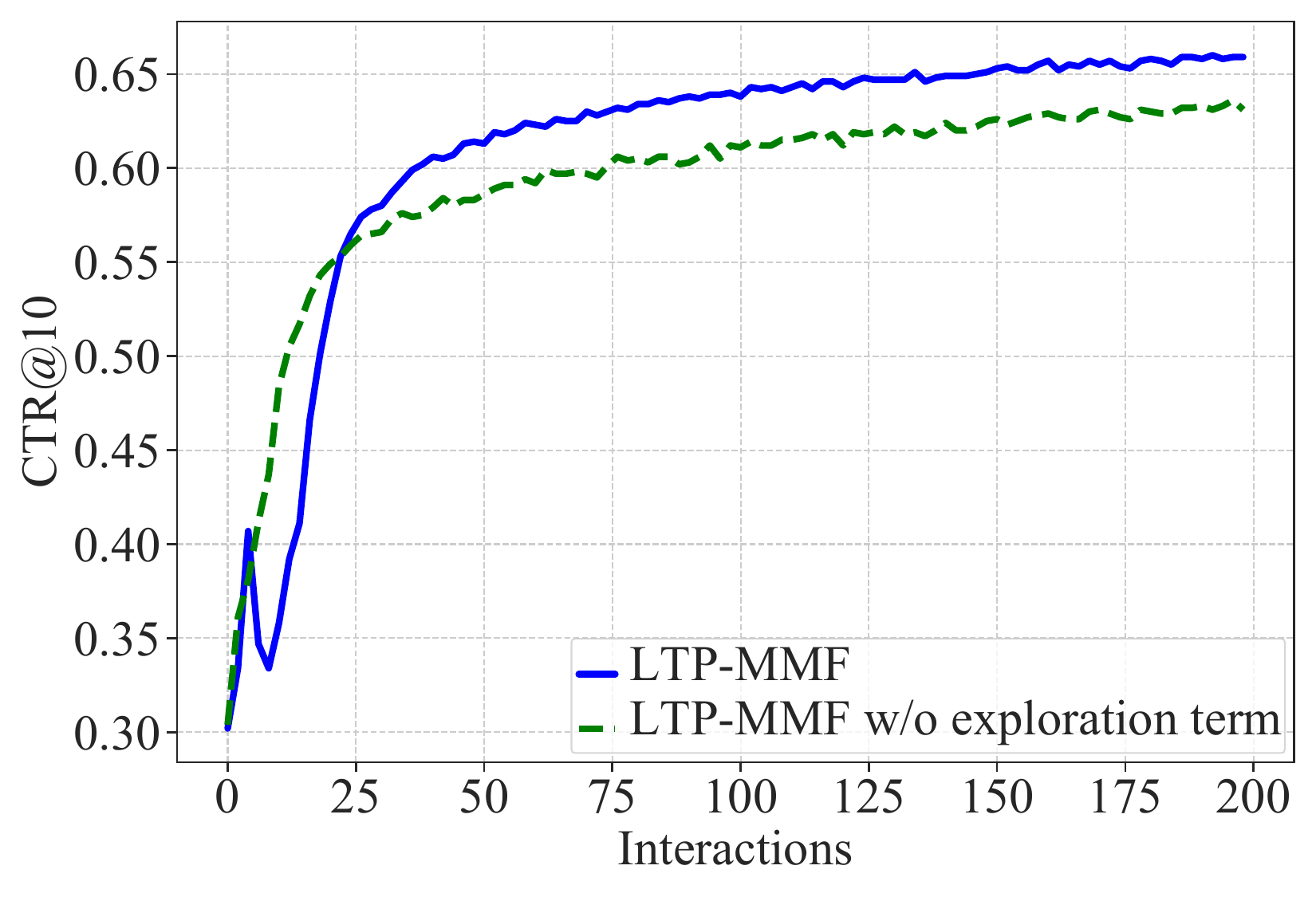}
\label{fig:ablation_CTR_explore}
}
    \subfigure[MMF]{
        \centering
        \includegraphics[width=0.48\linewidth]{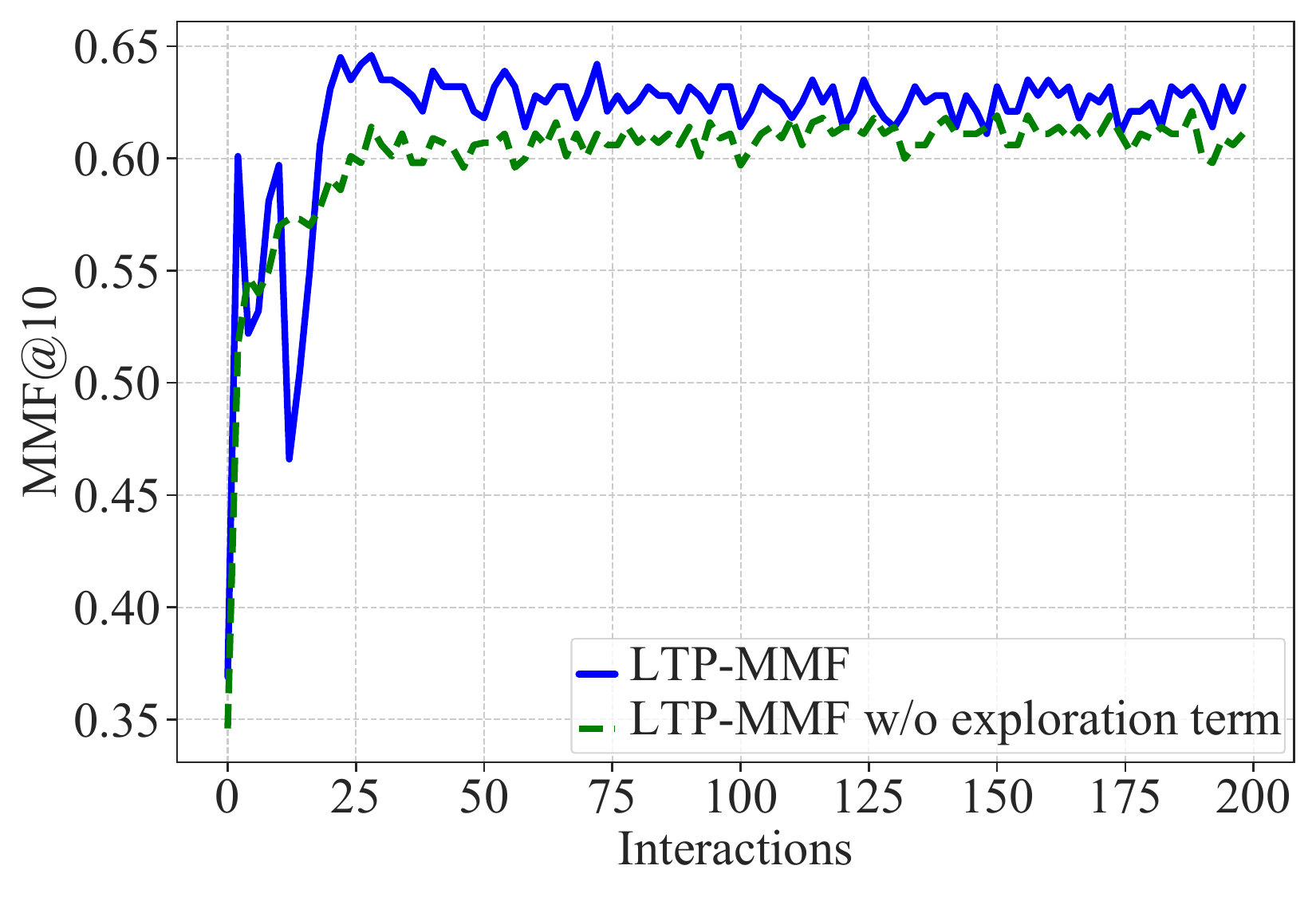}
        
        \label{fig:ablation_CTR_explore2}
}
    \caption{Ablation study for the exploration of LTP-MMF.}
    \label{fig:explore}
\end{figure}

\subsubsection{Overall performance}
Table~\ref{tab:EXP:main} reports the experimental results of LTP-MMF and the baselines on all three datasets to investigate the long-term performance. Underlined numbers mean the best-performed baseline and $*$ means the improvements over the best baseline are statistically significant ($t$-test, $p$-value< 0.05). To make fair comparisons, all the baselines were tuned and used $r@K$ as the evaluation metric. Note that similar experiment phenomena have also been observed on other $\lambda$ values.

From the reported results, we found that LTP-MMF outperformed all of the bandit baselines including hLinUCB, EXP3, and BLTS, which verified that LTP-MMF can serve the provider fairness well in the long term. We also observed that LTP-MMF outperformed all the MMF-based baselines, indicating that LTP-MMF can better remedy the bias in the RFL compared to other provider-fair baselines. The reason is that LTP-MMF can explore the feedback of unexposed items, avoiding always treating unexposed items as negative ones.

Due to exposure bias, larger exposure sizes tend to decrease bias, as they offer more items to be exposed to users. However, in modern Recommender Systems, such as YouTube~\cite{youtubeDNN}, the ranking size exposed to users is typically limited to single-digit items. Moreover, in the example depicted in Figure~\ref{fig:intro}, we can observe that as the ranking size increases, the severity of exposure bias diminishes. From the empirical studies presented in Table 2, we can observe that the average improvements over the best baselines are 5.5\%, 2.7\%, and 1.8\% for ranking sizes of 5, 10, and 20, respectively. This means that the smaller the magnitude of the exposure size (i.e. the larger the exposure bias in RFL), the more improvements LTP-MMF increases. It indicates the effectiveness of LTP-MMF especially when the ranking size is small.

Finally, it is evident that our model LTP-MMF significantly outperforms the long-term fairness baseline Dyn, underscoring the effectiveness of our methods in optimizing long-term fairness amidst the changing user preference estimations influenced by the feedback loops of RS.

\subsubsection{Pareto frontier}
Figure~\ref{fig:pareto} shows the Pareto frontiers~\cite{lotov2008visualizing} of CTR@K and MMF@K. The Pareto frontiers were drawn by tuning the accuracy-fairness trade-off coefficient $\lambda\in[1e-3,1]$ with best (CTR@K, MMF@K) long-term performances. In the experiment, we selected the baselines of P-MMF, CPFair, fairco, Welf, FairRec+, and BLTS, which achieved relatively good performances among fairness baselines and bandit baselines.

From the Pareto frontiers, we can see that the proposed LTP-MMF Pareto dominated all the baselines (i.e., the LTP-MMF curves are at the upper right corner) expect P-MMF~\cite{P-MMF} on the Amazon-Baby and Amazon-Beauty, indicating that LTP-MMF can achieve better user accuracy (i.e., CTR@K) with the same provider fairness (MMF@K) level. The results demonstrate that LTP-MMF can better trade-off accuracy and fairness in the long term.

Moreover, we can observe the changes in user accuracy and provider fairness with trade-off co-efficient $\lambda=[0.001,0.01]$. From Figure~\ref{fig:pareto_steam}, we can observe that when $\lambda$ increases from $0.001$ to $0.01$, the user accuracy (i.e., CTR@K) only decreases 0.09\% but provider fairness (i.e. MMF@K) increases 220\%. The result indicated that we can achieve better MMF in the long term while sacrificing little accuracy in LTP-MMF.

\begin{figure}
    \centering
\subfigure[Performance]{
        \centering
    \includegraphics[width=0.48\linewidth]{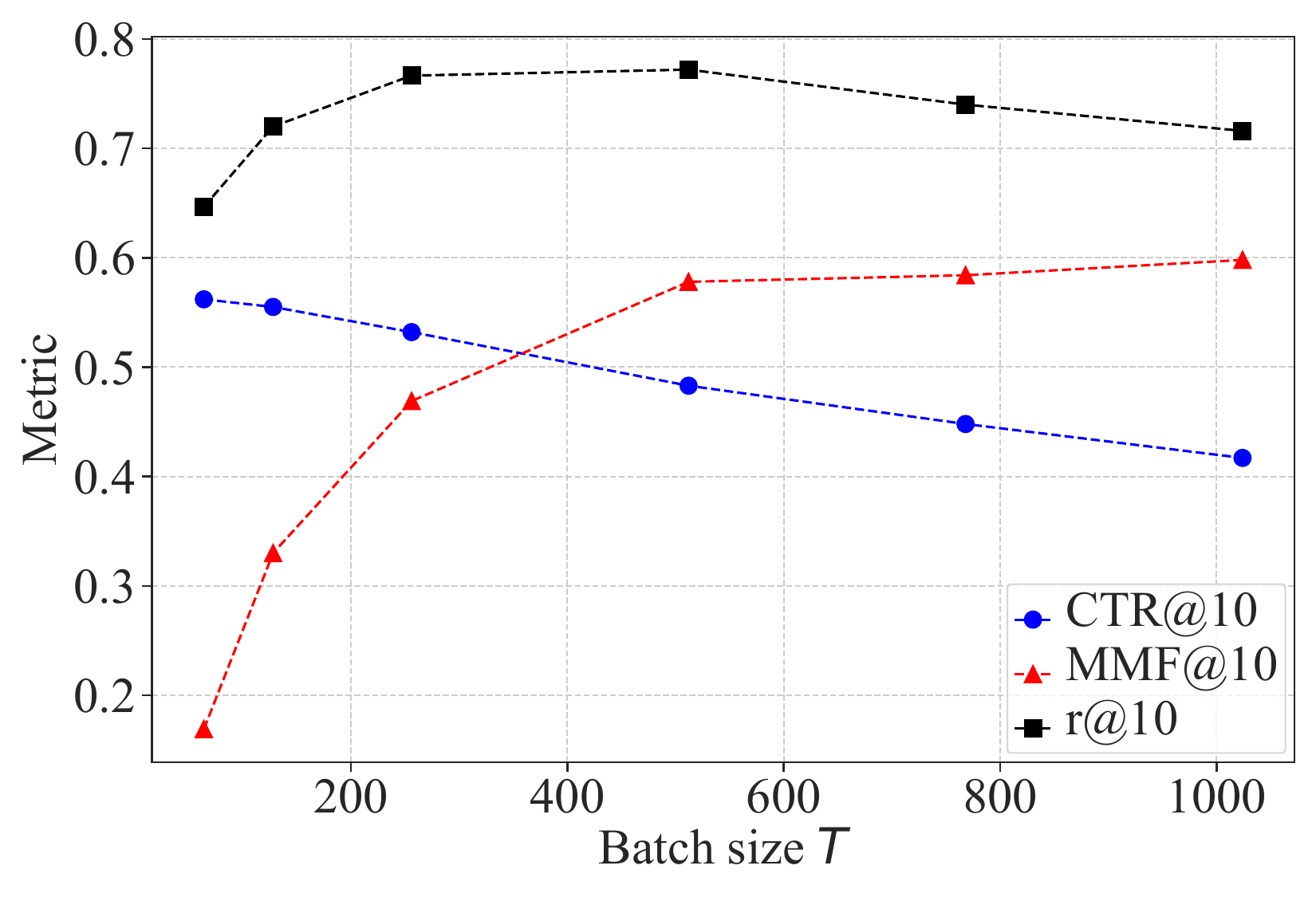}

        \label{fig:ablation_performance}
}
    \subfigure[Regret]{

        \centering
        \includegraphics[width=0.48\linewidth]{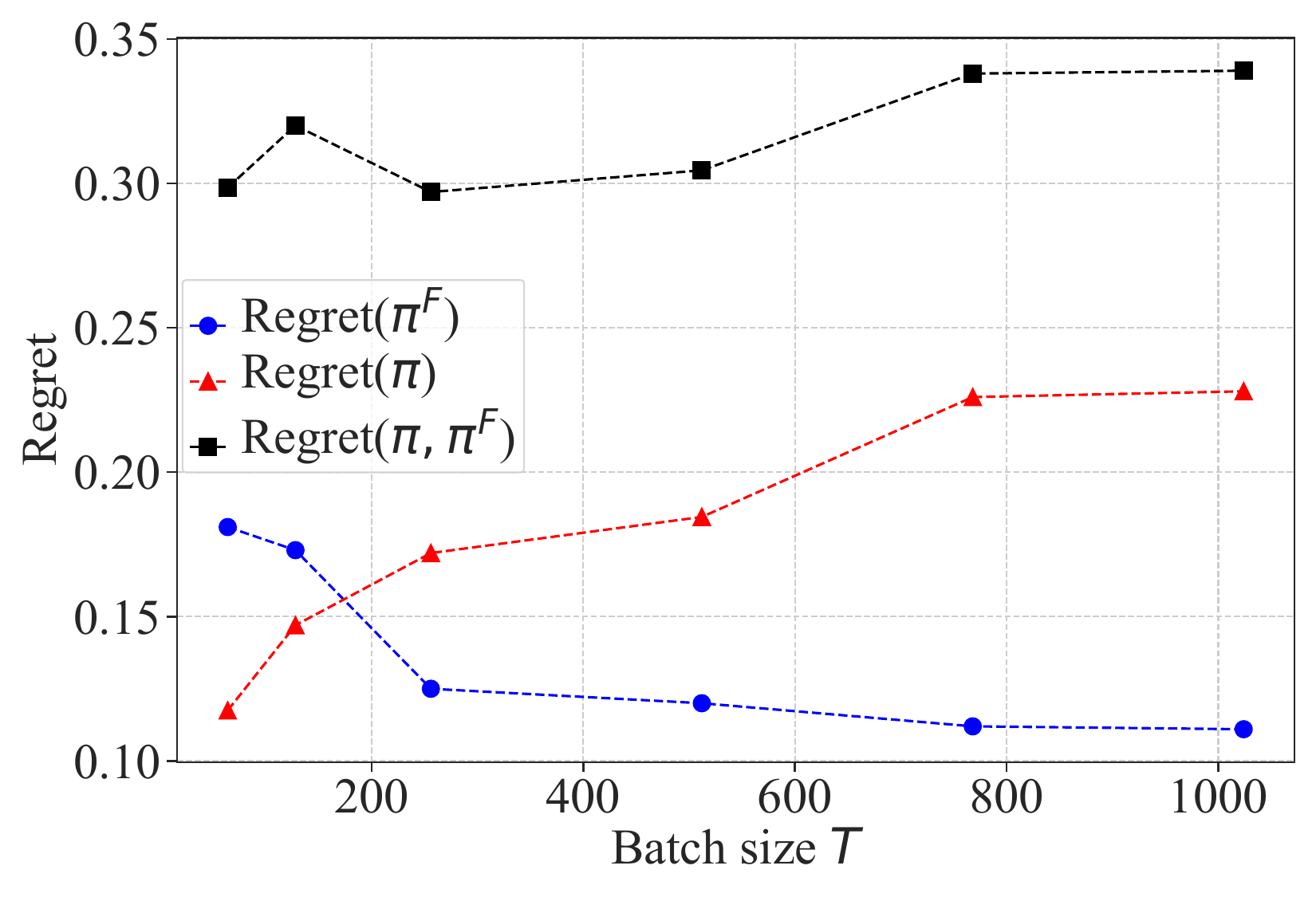}
      
        \label{fig:ablation_regret2}
   }
    \caption{(a) illustrates the long-term performance and (b) illustrates the long-term regret w.r.t. batch size $T$}
    \label{fig:regret}
\end{figure}

\begin{figure}
    \centering
    \includegraphics[width=0.9\linewidth]{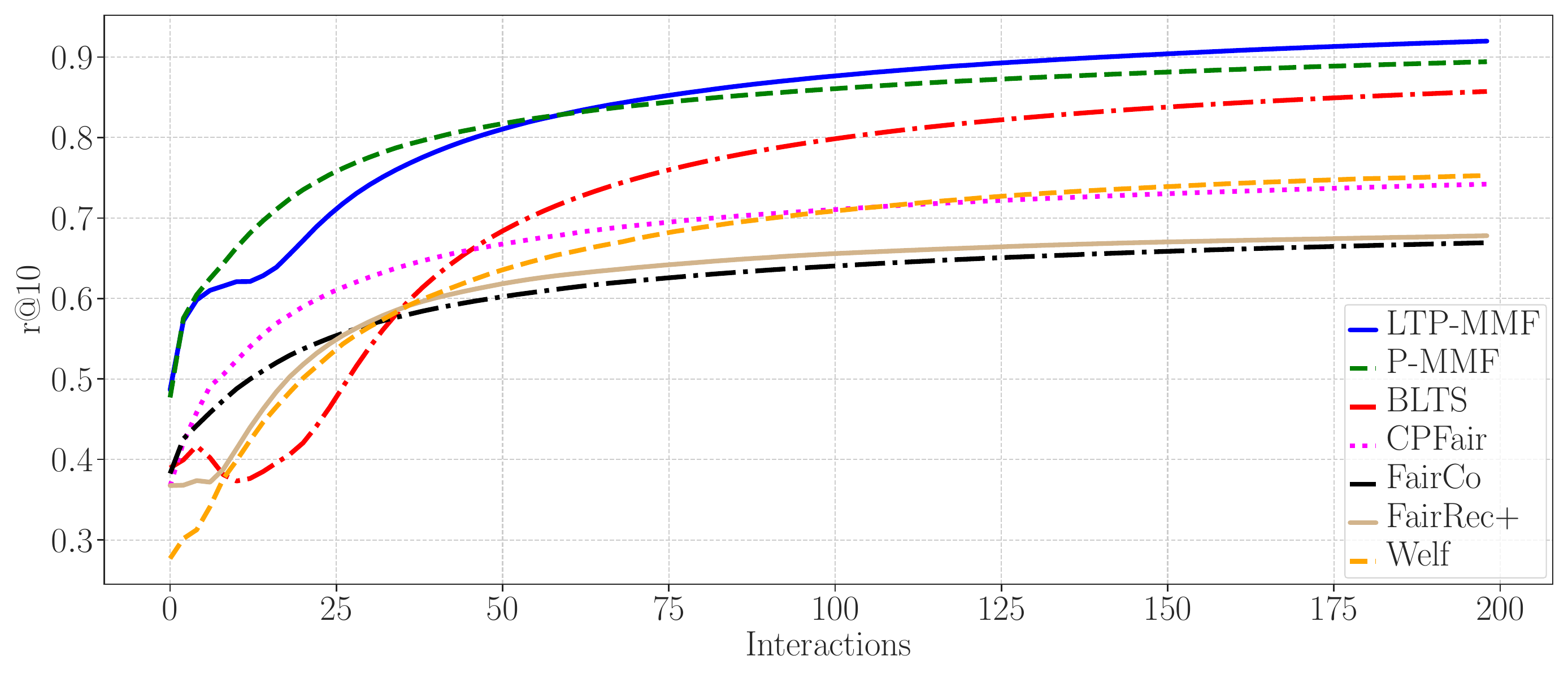}
    \caption{The periodical performance of LTP-MMF and other top baselines}
    \label{fig:steam_ablation_lt_baseline}
\end{figure}

\begin{figure}
    \centering
    \includegraphics[width=0.9\linewidth]{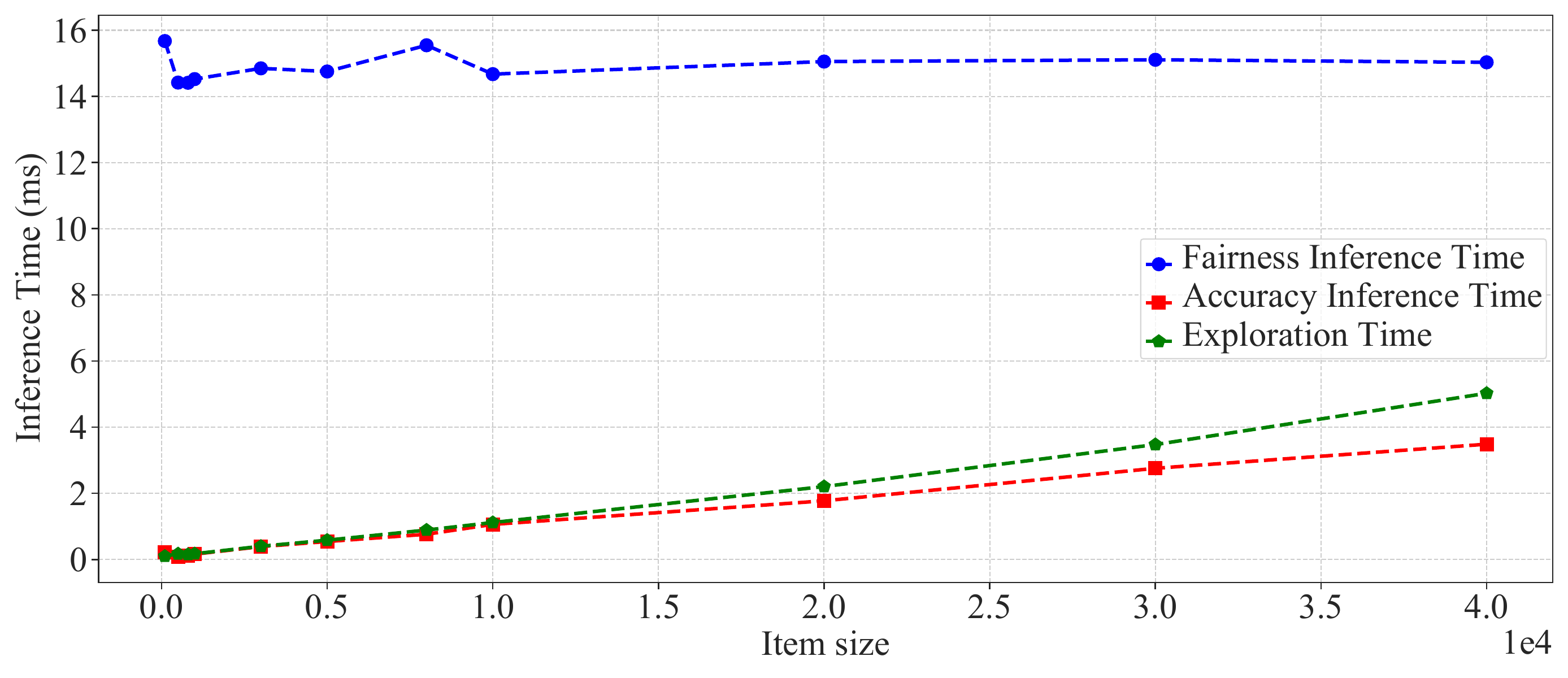}
    \caption{The online inference time per user w.r.t. the number of total items.}
    \label{fig:inference_time}
\end{figure}

\begin{figure}
        \centering
\includegraphics[width=0.7\linewidth]{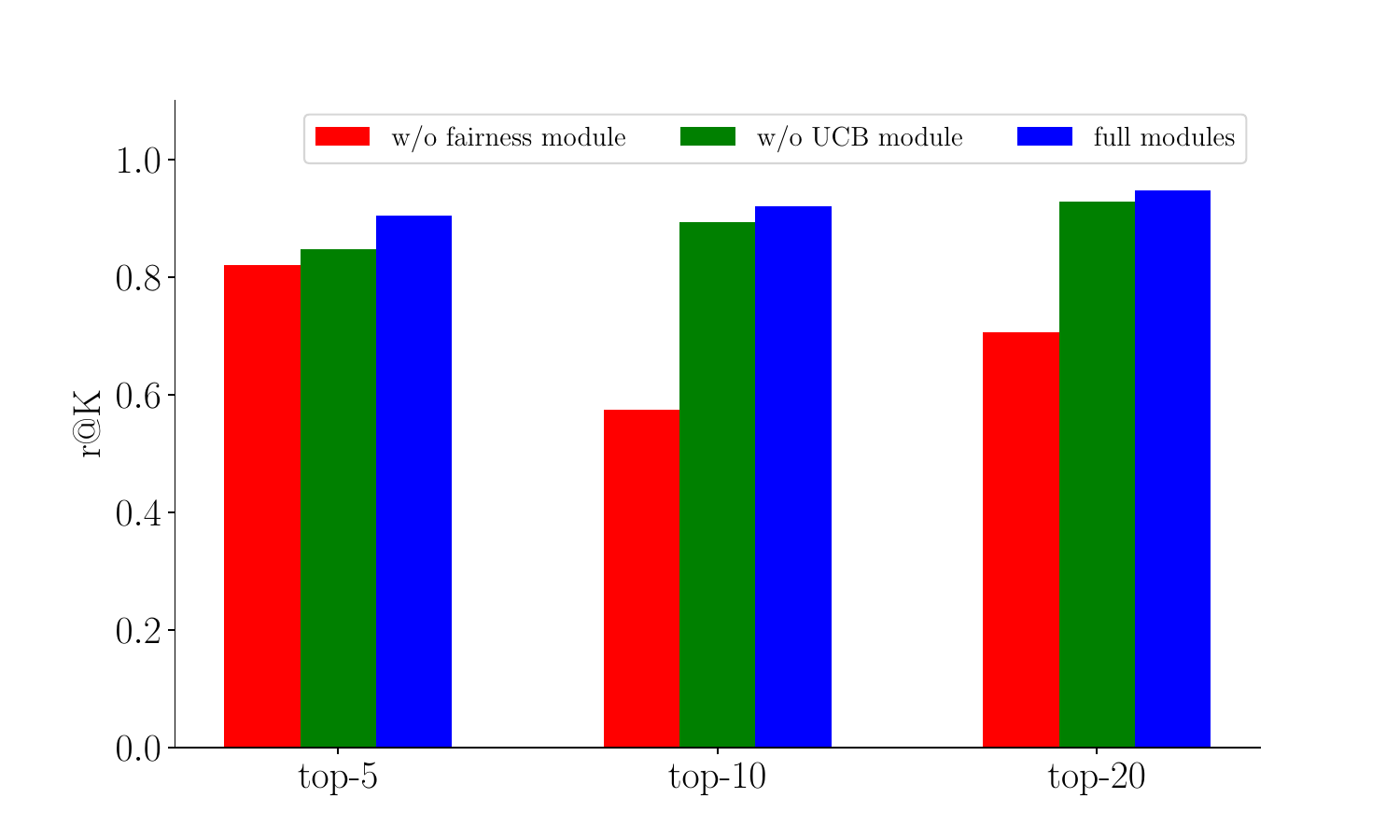}
        \caption{Ablation studies of the individual impact of each component. We report the overall performance $r@K$ on Steam datasets under different ranking sizes $K$. }\label{fig:ablation_module} 
\end{figure}

\subsection{Experiment Analysis}
We also conducted experiments to analyze LTP-MMF on the Steam dataset under top-$10$ settings.

\subsubsection{Ablation study on exploration term}\label{sec:ab4BM}
To investigate the impact of exploration term (i.e., the upper-confidence-bound $\triangle f_{u,i}$ in Eq.~\eqref{eq:cr4rankning}), we conduct an ablation study shown in Figure~\ref{fig:explore}. Specifically, we showed the CTR@10 and MMF@10 of two LTP-MMF variations, including the variations without exploration term (denoted as ``LTP-MMF w/o exploration term''), and the complete LTP-MMF. 
From Figure~\ref{fig:explore}, we found that in the first 25 interactions, the exploration term of LTP-MMF is dominant, leading to relatively poor performance. However, in the rest of the periods, LTP-MMF explores more trustful feedback of items, leading to better estimation of user preference. Therefore, LTP-MMF has better performance in terms of both user accuracy and provider fairness in the long term compared to ``LTP-MMF w/o exploration term''. The experiment also verified the importance of giving the exposures fairly to the providers in the feedback loop.

At the same time, we also conduct the experiment to investigate the long-term impact of different exploration weights (i.e. $\sqrt{\lambda_i}, \sqrt{\lambda_u}$ in Eq.~\ref{eq:bias_term}) shown in Figure~\ref{fig:alpha_u}. From the curve, we find that the performance improved when $\sqrt{\lambda_i}, \sqrt{\lambda_u} \in[0.01,1]$ and then they dropped between $[1,100]$, respectively. The reason is that the more exploration, the less bias will be in the long term. However, too much exploration will inevitably hurt the exploit reward, hurting the long-term performance. The opposite also holds. Therefore, we need to balance the exploration-exploit in the long term.

\subsubsection{Ablation study of different fair-aware modules}
To investigate the impact of different fair-aware modules (fairness module and UCB module),  we conduct an ablation study shown in Figure~\ref{fig:ablation_module}. Specifically, we showed the r@K of two LTP-MMF variations, including the variations without fairness module (denoted as ``w/o fairness module''), without UCB module (denoted as ``w/o UCB module''), and the complete LTP-MMF (denoted as ``full modules''). 

From Figure~\ref{fig:ablation_module}, We can observe that when dropping out of the fairness and UCB modules, the overall performance decreases under different ranking sizes, indicating the effectiveness of the different fairness-aware modules.

\subsubsection{Impact of the batch size $T$}\label{sec:batch_size}
According to Theorem~\ref{theo:regret}, the batch size $T$ balances the regret of user accuracy and provider fairness. In this experiment, Figure~\ref{fig:regret} (a) studied how CTR@K, MMF@K, and r@K changed when the batch size $T$ was set to different values from $[64,1024]$. Figure (b) studied how regret of accuracy (Regret($\pi$)), regret of fairness (Regret($\pi^F$)) and regret of LTP-MMF (Regret($\pi,\pi^F$)) changed when the batch size $T$ was set from $[64,1024]$. 

From the curves shown in Figure~\ref{fig:regret} (a), we found the accuracy performance CTR@K improved while the fairness performance MMF@K dropped when the batch size becomes smaller. Similarly, in Figure~\ref{fig:regret} (b), the regret of accuracy Regret($\pi$) dropped while the regret of fairness Regret($\pi^F$) dropped when the batch size becomes smaller. The results verified the theoretical analysis that small $T$ (e.g., $T=64$) results in more bias in fairness estimation while large $T$ (e.g., $T=1024$) results in large bias in accuracy estimation in the long term.

Moreover, The overall performance $r@K$ improved and regret of LTP-MMF (Regret($\pi,\pi^F$)) dropped when $T\in[64,512]$ and then $r@K$, Regret($\pi,\pi^F$) dropped and improved between $[512,1024]$, respectively. Therefore, it is important to balance the accuracy and fairness in real applications through batch size $T$.

\subsubsection{Periodical performance}
Figure~\ref{fig:steam_ablation_lt_baseline} reports the experimental results of LTP-MMF and performing baselines to investigate how the LTP-MMF performs for each period. Note that similar experiment phenomena have also been observed on other $\lambda$ and top-$k$ values. To make fair comparisons, all the baselines were tuned $\lambda=1$ as the evaluation metric.

From the $r@10$ curve, we found that although the performance gaps between LTP-MMF and other baselines are not obvious in the beginning (interaction 0), the LTP-MMF can well explore the feedback of unexposed items, leading to a huge improvement over other baselines in the long term (interaction 200). Moreover, in most periods, LTP-MMF can steadily outperform other baselines, verifying the effectiveness of LTP-MMF. 

\subsubsection{Online inference time}

We experimented with investigating the online inference time of LTP-MMF. 
Figure~\ref{fig:inference_time} reports the curves of inference time (ms) of the accuracy module, fairness module, and exploration term computation per user access w.r.t. item size. We can see that LTP-MMF with GPU versions needs only within 6ms to calculate user-item scores $s_{u,i}$ and exploration term $\triangle f_{u,i}$. The reason is that this operation only needs matrix multiplication and the inverse of the matrix can be stored in the training (offline) phase, leading to low latency in the online phase. Moreover, the inference time for the fairness module can also be maintained as 14-16ms even when the item size becomes larger. We conclude that LTP-MMF can be adapted to online scenarios efficiently because of its low latency, even when the item size grows rapidly.

\section{Conclusion}
We proposed a novel ranking model called LTP-MMF that aims to consider provider max-min fairness in the long term. Firstly, we formulated the provider fair recommendation as a repeated resource allocation problem under a batched bandit setting. LTP-MMF applies the exploration term to break the loop while exploiting the fairness-aware rewards. Our theoretical analysis showed that the regret of LTP-MMF can be bounded. Experiments on four available datasets demonstrated that LTP-MMF can conduct ranking in an effective and efficient way.

\begin{acks}
This work was funded by the National Key R\&D Program of China (2023YFA1008704), the National Natural Science Foundation of China (No. 62376275 \& No. 62106273 \& No. 72192805), Engineering Research Center of Next-Generation Intelligent Search and Recommendation, Ministry of Education, Major Innovation \& Planning Interdisciplinary  Platform for the ``Double-First Class'' Initiative, Renmin University of China, fund for building world-class universities (disciplines) of Renmin University of China. Supported by the Outstanding Innovative Talents Cultivation Funded Programs 2024 of Renmin University of China.
\end{acks}


\newpage
\appendix
    \section{Appendix}

\subsection{Proof of Theorem 1}\label{app:dual_prove}
\begin{proof}
For max-min fairness, we have the regularizer as
$
    r(\mathbf{e}) = \min_{p\in\mathcal{P}} \left(\mathbf{e}_p/\bm{\gamma}_p\right),
$
we can easily proof that the exposure vector $\mathbf{e}$ can be represented as the dot-product between decision varible $\mathbf{x}_t$ and the item-provider adjacent matrix $\mathbf{A}$:
$
    \mathbf{e} = \sum_{t=1}^T\left(\mathbf{A}^{\top}\mathbf{x}_t\right).
$
Then we treat the $\mathbf{e}$ as the auxiliary variable, and the ideal objective can be written as:
\begin{align*}
    W_{OPT} &= \max_{\mathbf{x}_t\in\mathcal{X},\mathbf{e}\leq \bm{\gamma}}\left[\sum_{t=1}^Tg(\mathbf{x}_t)/T + \lambda r(\mathbf{e})\right]\\
    s.t. \mathbf{e} &= \sum_{t=1}^T\left(\mathbf{A}^{\top}\mathbf{x}_t\right),
\end{align*}
where $\mathcal{X} = \{\mathbf{x}_t|\mathbf{x}_t \in {0,1} \land \sum_{i\in\mathcal{I}} \mathbf{x}_{ti} = K\}$. 
Then we move the constraints to the objective using a vector of Lagrange multipliers $\boldsymbol{\mu}\in\mathbb{R}^{|\mathcal{P}}|$:
\begin{align*}
    &W_{OPT} = \max_{\mathbf{x}_t\in\mathcal{X},\mathbf{e}\leq \bm{\gamma}}\min_{\boldsymbol{\mu}\in\mathcal{D}}\left[\sum_{t=1}^Tg(\mathbf{x}_t)/T + \lambda r(\mathbf{e}) - \boldsymbol{\mu}^{\top} \left(-\mathbf{e}+\sum_{t=1}^T\mathbf{A}^{\top}\mathbf{x}_t\right)\right]\\
    & \leq \min_{\boldsymbol{\mu}\in\mathcal{D}}\left[\max_{\mathbf{x}_t\in\mathcal{X}}\left[\sum_{t=1}^Tg(\mathbf{x}_t)/T - \boldsymbol{\mu}^{\top}\sum_{t=1}^T\mathbf{A}^{\top}\mathbf{x}_t\right] + \max_{\mathbf{e}\leq \bm{\gamma}}\left(\lambda r(\mathbf{e})-\boldsymbol{\mu}^{\top}\mathbf{e}\right)\right]\\
    &=\min_{\boldsymbol{\mu}\in\mathcal{D}}\left[f^*(\mathbf{A}\boldsymbol{\mu}) + \lambda r^{*}(-\boldsymbol{\mu})\right] = W_{Dual},
\end{align*}
where $\mathcal{D} = \boldsymbol{\mu}|r^*(-\boldsymbol{\mu})<\infty\}$ is the feasible region of dual variable $\boldsymbol{\mu}$. According to the Lemma 1 in the ~\citet{balseiro2021regularized}, we have $\mathcal{D}$ is convex and positive orthant is inside the recession cone of $\mathcal{D}$.




We let the variable $\mathbf{z}_p = (\mathbf{e}_p/\bm{\gamma}_p-1)$, we have:
\begin{align*}
    r^*(\boldsymbol{\mu}) &= \max_{\mathbf{e}\leq\bm{\gamma}}\left[\min\left(\mathbf{e}_p/\bm{\gamma}_p\right) + \boldsymbol{\mu}^{\top}\mathbf{\mathbf{e}}/\lambda\right]\\
    &= \boldsymbol{\mu}^{\top}\bm{\gamma} /\lambda +  1 + \max_{\mathbf{z}_p \leq 0}\left[\min(\mathbf{z}_p) + 1/\lambda \sum_{p\in\mathcal{P}}\boldsymbol{\mu}_p\bm{\gamma}_p\mathbf{z}_p\right]
\end{align*}
    
Let $s(\mathbf{z}) = \min_p \mathbf{z}_p$ and $\mathbf{v} = (\boldsymbol{\mu}\odot\bm{\gamma})/\lambda$, $\odot$ is the hadamard product. Then we define $s^*(\mathbf{v}) = \max_{\mathbf{z}\leq 0}\left(s(\mathbf{z}) + \mathbf{z}^T\mathbf{v}\right)$. We firstly show that if $\sum_{p\in\mathcal{S}}\mathbf{v}_p \ge -1, \forall \mathcal{S}\in\mathcal{P}_s$, then $s^*(\mathbf{v}) = 0$ and $\mathbf{z} = 0$ is the optimal solution, otherwise $s^*(\mathbf{v}) = \infty$.

We can equivalently write $\mathcal{D} = \{\mathbf{v}|\sum_{p\in\mathcal{S}}\mathbf{v}_p \ge -1, \forall \mathcal{S}\in\mathcal{P}_s\}$. We firstly show that $s^*(\mathbf{v}) = \infty$ for $\mathbf{v}\notin \mathcal{D}$. Suppose that there exists a subset $\mathcal{S}\in\mathcal{P}_s$ such that $\sum_{p\in\mathcal{S}} \mathbf{v}_p < -1$. For any $b > 1$, we can get a feasible solution:
\begin{align*}
\begin{split}
\mathbf{v}_p= \left \{
\begin{array}{ll}
   -b,                    & p\in \mathcal{S}\\
    0,                    & otherwise.
\end{array}
\right.
\end{split}
\end{align*}
Then, because such solution is feasible and $s(\mathbf{z}) = -b$, we obtain that 

\[s^*(\mathbf{v}) \ge s(\mathbf{z}) - b(\sum_{p\in\mathcal{S}}\mathbf{v}_p) = -b(\sum_{p\in\mathcal{S}}\mathbf{v}_p+1).
\]
Let $b\rightarrow\infty$, we have $s^*(\mathbf{v})\rightarrow\infty$.

Then we show that $s^*(\boldsymbol{\mu}) = 0$ for $\mathbf{v}\in\mathcal{D}$. Note that $\mathbf{z} = 0$ is feasible. Therefore, we have
\[
    s^*(\mathbf{v})\ge s^*(0) = 0.
\]
Then we have $\mathbf{z} \leq 0$ and without loss of generality, that the vector $\mathbf{z}$ is sorted in increasing order, i.e., $\mathbf{z}_1\leq \mathbf{z}_2, \cdots, \leq \mathbf{z}_{|\mathcal{P}|}$.
The objective value is
\begin{align*}
     s^*(\mathbf{v}) &= \mathbf{z}_1 + \sum_{j\in|\mathcal{P}|}\mathbf{z}_j\mathbf{v}_j \\
     &= \sum_{j=1}^{|\mathcal{P}|}\left(\mathbf{z}_j-\mathbf{z}_{j+1}\right)\left(1+\sum_{i=1}^{j}\mathbf{v}_j\right)\leq 0.
\end{align*}
\end{proof}

\subsection{Proof of Lemma 1}
\begin{proof}
From the proof of Theorem 1, we have 
\[
s^*(\bm{\mu}) = \max_{\bm{\mu}\leq 0}(\min_p \bm{z}_p + \bm{z}^{\top}\bm{\mu})
\]
and $s^*(\boldsymbol{\mu}) = 0$ for $\mathbf{v}\in\mathcal{D}$ . Therefore, we have 
\[
    r^*(-\boldsymbol{\mu}) = \boldsymbol{\mu}^{\top}\bm{\gamma}/\lambda + 1.
\]

\end{proof}

\subsection{Proof of Theorem 2}\label{app:mf_ucb_prove}
\begin{proof}

        Our proof will take the following four steps:
        
        \textbf{Bias Term Bound} Firstly, we will bound the bias term $\|\hat{\bm{v}}_{u,t}-\bm{v}_u^*\|_{\bm{A}_{u,t}}, \|\hat{\bm{v}}_{i,t}-\bm{v}_i^*\|_{\bm{C}_{i,t}}$:

        we define the bias term upper bound as $\alpha_{t}, \beta_{t}$ respectively. We will bound the bias term as follows:
        by taking the gradient of the objective function  respect to $v_{u},v_{i}$, we have,
        \begin{equation}
            \bm{A}_{u,t}(\hat{\bm{v}}_{u,t}-\bm{v}_u^*) = \sum_{j=1}^t \hat{\bm{v}}_{i,j}(\bm{v}_i^*-\hat{\bm{v}}_{i,j})^{\top}\bm{v}_u^* + \sum_{j=1}^t (\hat{\bm{v}}_{i,j})\epsilon_j - \lambda_u \bm{v}_u^*,
        \end{equation}
        where $\epsilon_j$ is the Gaussian noise at time $j$. 
        Without loss of generality, in ranking task, we can always scale the user-item score $s_{u,i}$, so that the l2-norm of $v_u,v_i$ can be bounded by a constant factor:
        \[
            \|v_u\|_2 \leq 1, \quad \|v_i\|_2 \leq 1
        \]
        Therefore, we can bound the function norm of the above equation as
        \begin{equation}
        \begin{aligned}
            \|\hat{\bm{v}}_{u,t}-v_u^*\|_{\bm{A}_{u,t}} &= \|\sum_{j=1}^t \hat{\bm{v}}_{i,j}(v_i^*-\hat{\bm{v}}_{i,j})^{\top}\bm{v}_u^* + \sum_{j=1}^t (\hat{\bm{v}}_{i,j})\epsilon_j - \lambda_u \bm{v}_u^*\|_{\bm{A}_{u,t}} \\
            & \leq \|\sum_{j=1}^t \hat{\bm{v}}_{i,j}\epsilon_j\|_{\bm{A}_{u,t}} + \frac{1}{\sqrt{\lambda_u}}\sum_{j=1}^t\|\bm{v}_{i}^*-\hat{\bm{v}}_{i,j}\|_2 + \sqrt{\lambda_u}
        \end{aligned}
        \end{equation}
         Since the $q$-linearly convergent to the optimizer of parameter $\bm{v}_u,\bm{v}_i$ in~\cite{wang2016learning}, we have for every $\epsilon_q>0$ and $0 < q < 1$, we have
        \begin{equation}\label{eq:q-converge}
            \|\bm{v}_{i}^*-\hat{\bm{v}}_{i,t+1}\|_2 \leq  (q+\epsilon_q)\|\bm{v}_{i}^*-\hat{\bm{v}}_{i,t}\|_2
        \end{equation}
            
         Therefore, by applying the self-normalized vector-valued martingales~\cite{abbasi2011improved}, we have for any $\sigma > 0$, with probability at least $1-\sigma$, 
         \[
         \|\hat{\bm{v}}_{u,t}-v_u^*\|_{\bm{A}_{u,t}} \leq \alpha_{t},
         \]
         \[
            \alpha_{t} = \sqrt{\lambda_u} + \frac{2(q+\epsilon_q)(1-(q+\epsilon_q)^t)}{1-q-\epsilon_q} + \sqrt{d\ln{\frac{\lambda_ud+t}{\lambda_ud\sigma}}}.
         \]
         Similarly, we have
         \[
         \|\hat{\bm{v}}_{i,t}-v_i^*\|_{\bm{C}_{i,t}} \leq \beta_{t},
         \]
         \[
            \beta_{t} = \sqrt{\lambda_i} + \frac{2(q+\epsilon_q)(1-(q+\epsilon_q)^t)}{1-q-\epsilon_q} + \sqrt{d\ln{\frac{\lambda_id+t}{\lambda_id\sigma}}}
         \].
        Therefore, the bias term $\alpha_{t},\beta_{t}$ are comparable with $O((1-(q+\epsilon_r)^t)\sqrt{\ln t})$ for every $\epsilon_r>0$.

        \textbf{Variance Term Bound} Next, we will bound the variance term 
        Then we prove the converge rate of variance term: $ \|\hat{\bm{v}}_{u,t}\|^2_{\bm{A}_{u,t}}, \|\hat{\bm{v}}_{i,t}\|^2_{\bm{C}_{i,t}}$ as follows:
        \begin{equation}
             \begin{aligned}
            \|\hat{\bm{v}}_{u,t}\|^2_{\bm{A}_{u,t}} &= \hat{\bm{v}}_{u,t}^{\top}\bm{A}_{u,t}\hat{\bm{v}}_{u,t}\\
            &= \hat{\bm{v}}_{u,t}^{\top}(\lambda_u \bm{I} + \sum_{j=1}^t \hat{\bm{v}}_{i,t}\hat{\bm{v}}_{i,t}^{\top})\hat{\bm{v}}_{u,t}\\
            &\leq \|\hat{\bm{u}}_{i,t}\|_2^2 + \sum_{j=1}^T\|\hat{\bm{v}}_{i,j}^{\top}\hat{\bm{v}}_{u,t}\|_2^2 \leq (t+1) \sim O(t),
            \end{aligned}  
        \end{equation}
       
        since we can always scale the l2-norm of $v_u,v_i$ by any rate in ranking tasks (we only care the relative score). Thus we can easily obtain that
        \[
        \|\hat{\bm{v}}_{u,t}\|_{\bm{A}_{u,t}^{-1}} \sim \O(1/\sqrt{t}).
        \]

        \textbf{Collaborative Variance Term Bound} Next, we will bound the variance term of collaborative term $ \|\hat{\bm{v}}_{u,t}-\bm{v}_u^*\|^2_{\bm{A}^{-1}_{u,t}}, \|\hat{\bm{v}}_{i,t}-\bm{v}_i^*\|^2_{\bm{C}^{-1}_{i,t}}$ as follows:
        \begin{align*}
            \|\hat{\bm{v}}_{u,t}-\bm{v}_u^*\|_{\bm{A}^{-1}_{u,t}} \leq \|\hat{\bm{v}}_{u,t}-\bm{v}_u^*\|_2 \leq (q+\epsilon_q)^t,
        \end{align*}
        where $q, \epsilon_q$ follows the Eq.~\eqref{eq:q-converge}.
        Similarly, we have 
        \begin{align*}
            \|\hat{\bm{v}}_{i,t}-\bm{v}_i^*\|_{\bm{C}^{-1}_{i,t}} \leq \|\hat{\bm{v}}_{i,t}-\bm{v}_i^*\|_2 \leq (q+\epsilon_q)^t.
        \end{align*}
        Let's abbreviate the upper bound of collaborative error term $(q+\epsilon_q)^t$ as $C$

        \textbf{Upper Confidence Bound}
        Finally, the upper confidence bound of user-item score can have easily by putting the aforementioned term together:
        \begin{equation}
             \begin{aligned}
                &s_{u,i}^* - \hat{s}_{u_t,i} = (\bm{v}_u^*)^{\top}\bm{v}_i^* - \hat{v}_{u,t}^{\top}\hat{v}_{i,t} \\
                &= 1/2\left[(\bm{v}_u^*+\hat{v}_{u,t})^{\top}(\bm{v}_i^*-\hat{v}_{i,t}) + (\bm{v}_u^*-\hat{v}_{u,t})^{\top}(\bm{v}_i^*+\hat{v}_{i,t})\right] \\
                &\leq \alpha_t(C/2+\|\hat{v}_{i,t}\|_{A_{u,t}^{-1}}) + \beta_t(C/2+\|\hat{v}_{u,t}\|_{C_{i,t}^{-1}}).
            \end{aligned}
        \end{equation}
        We can easily have the upper bound of user-item score have the converge term 
        \begin{equation}
            O(\frac{(1-(q+\epsilon_q)^t)\sqrt{\ln t}}{\sqrt{t}}), 
        \end{equation}
        which is a decrease function of $t$ when the $t$ becomes large.
    \end{proof}
    

\subsection{Proof of Theorem 3}
\begin{proof}
Firstly, in practice, we normalize the user-item preference score $s_{u,i}$ to $[0,1]$. Therefore, $\sum_{t=1}^Tg(\mathbf{x}_t)/T\leq K$. In max-min regularizer $r(\mathbf{e})$. Let's abbreviate its upper bound to $\bar{r}$. In practice, $\bar{r}\leq 1$ We have
\begin{equation}
    W_{OPT} \leq K + \lambda \bar{r}.
\end{equation}

We consider the stopping time $\tau$ of Algorithm~\ref{alg:bandit-MMF} as the first time the provider will have the maximum exposures, i.e.
\[
    \sum_{t=1}^{\tau}\mathbf{\bm{M}}^{\top}\mathbf{x}_t \ge \bm{\gamma}.
\]
Note that is $\tau$ a random variable.

Similarly, following the proven idea of~\citet{balseiro2021regularized}, first, we analyze the primal performance of the objective function. Second, we bound the complementary slackness term by the momentum gradient descent. Finally, We conclude by putting it to achieve the final regret bound.

\textbf{Primal performance proof}: Consider a time $t<\tau$, the recommender action will not violate the resource constraint. Therefore, we have:
\[
    g(\mathbf{x}_t)/T = g^*(\bm{M}\boldsymbol{\mu}_t) + \lambda {\boldsymbol{\mu}_t}^T\bm{M}^{\top}\mathbf{x}_t,
\]
and we have $\mathbf{e}_t = \argmax_{\mathbf{e}\leq \boldsymbol{\gamma}}\{r(\mathbf{e}) + \boldsymbol{\mu}^{\top}\mathbf{e}/\lambda\}$
\[
    r({\mathbf{e}_t}) = r^*(-\boldsymbol{\mu}) - \mathbf{\boldsymbol{\mu}_t}^T\mathbf{e}_t/\lambda.
\]

We make the expectations for the current time step $t$ for the primal functions:
\begin{align*}
    \mathbb{E}\left[g(\mathbf{x}_t)/T + \lambda  r({\mathbf{e}_t})\right] &= \mathbb{E}\left[g^*(\bm{M}\boldsymbol{\mu}_t) + \mathbf{\boldsymbol{\mu}_t}^T\bm{M}^{\top}\mathbf{x}_t + \lambda r^*(-\boldsymbol{\mu}) - \mathbf{\boldsymbol{\mu}_t}^T\mathbf{e}_t\right] \\
    &= W_{Dual}(\boldsymbol{\mu}_t) - \mathbf{E}\left[\boldsymbol{\mu}_t^T(-\bm{M}^{\top}\mathbf{x}_t + \mathbf{e}_t)\right].
\end{align*}

Consider the process $Z_t = \sum_{j=1}^T\boldsymbol{\mu}_j^t(-\bm{M}^{\top}\mathbf{x}_t + \mathbf{e}_t)-\mathbf{E}\left[\boldsymbol{\mu}_t^T(-\bm{M}^{\top}\mathbf{x}_t + \mathbf{e}_t)\right]$ is a martingale process. The Optional Stopping Theorem in martingale process~\cite{williams1991probability} implies that $\mathbb{E}\left[Z_{\tau}\right] = 0$. Consider the variable $w_t(\boldsymbol{\mu}_t) = \boldsymbol{\mu}_t^T(-\mathbf{A}^{\top}\mathbf{x}_t + \mathbf{e}_t)$, we have
\begin{align*}
    \mathbb{E}\left[\sum_{t=1}^{\tau}w_t(\boldsymbol{\mu}_t)\right] = \mathbb{E}\left[\sum_{t=1}^{\tau}\mathbb{E}\left[w_t(\boldsymbol{\mu}_t)\right]\right]
\end{align*}

Moreover, in MMF, the dual function $W_{Dual}$ is convex proofed in Theorem~\ref{theo:dual}, we have
\begin{equation}
\begin{aligned}
    \mathbb{E}\left[\sum_{t=1}^{\tau}g(\mathbf{x}_t)/T + \lambda r(\mathbf{e}_t)\right]  &= \mathbb{E}\left[\sum_{t=1}^{\tau}W_{Dual}(\boldsymbol{\mu}_t)\right] - \mathbb{E}\left[\sum_{t=1}^{\tau}w_t(\boldsymbol{\mu}_t)\right]\\
    &\leq \mathbb{E}\left[\tau W_{Dual}(\widetilde{\boldsymbol{\mu}_{\tau}})\right] - \mathbb{E}\left[\sum_{t=1}^{\tau}w_t(\boldsymbol{\mu}_t)\right],
\end{aligned}
\end{equation}
where $\widetilde{\boldsymbol{\mu}_{\tau}} =\sum_{t=1}^{\tau}\boldsymbol{\mu}_t/\tau$.

Next, we will bound the bias of the primal performance due to the estimation error of the ranking model.

From the proof of theorem 1, 
we can bound the $W_{Dual}(\bm{\mu}_t)$ as follows:
at iteration $n$, for any $\sigma > 0$, with probability at least $1-\sigma$,
\begin{align*}
    W_{Dual}(\bm{\mu}_t)-\hat{W}_{Dual}(\bm{\mu}_t) &= g^*(\bm{M}\bm{\mu}_t) + \lambda r^*(-\mu) - \hat{g}^*(\bm{M}\bm{\mu}_t) - \lambda r^*(-\mu)\\
    &= s_u^{\top}\bm{x}_t - (\bm{M}\bm{\mu_t})^{\top}\bm{x}_t - \bm{s}_u^{\top}\bm{x}_t - (\bm{M}\bm{\mu_t})^{\top}\bm{x}_t \leq \sum_{i,\bm{x}_{ti}=1} \triangle f_{u_t,i}^n,
\end{align*}


\textbf{Complementary slackness proof}
Then we aim to proof the complementary slackness $\sum_{t=1}^Tw_t(\boldsymbol{\mu}_t) - w_t(\boldsymbol{\mu})$ is bounded. Suppose there exists $G, s.t.$ the gradient norm is bounded $
\| \widetilde{\mathbf{g}}_t \| \leq G $. Then we have:
\begin{equation}\label{eq:regret_mu}
    \sum_{t=1}^{\tau}w_t(\boldsymbol{\mu}_t) - w_t(\boldsymbol{\mu}) \leq \frac{L^2}{\eta} + \frac{G^2}{(1-\alpha)\sigma}\eta(\tau-1) + \frac{G^2}{2(1-\alpha)^2\sigma\eta},
\end{equation}
where the project function $\| \boldsymbol{\mu}-\boldsymbol{\mu}_t \|_{\boldsymbol{\gamma}}^2$ is $\sigma-$strongly convex.

Next, we prove the inequality in Equation. 
According to the Theorem 1 in ~\cite{momentum4online}, 
we have 
\[
    \| \mathbf{g}_t \|_2^2 = \|(1-\alpha)\sum_{s=1}^t\alpha^{t-s}(\widetilde{\mathbf{g}}_s)\|_2^2 \leq G^2,
\]
and 
\[
    \sum_{t=1}^{\tau}w_t(\boldsymbol{\mu}_t) - w_t(\boldsymbol{\mu}) \leq \frac{\|\boldsymbol{\mu}_t-\boldsymbol{\mu}_0\|_{\boldsymbol{\gamma}^2}^2}{\eta} + \frac{G^2}{(1-\alpha)\sigma}\eta(\tau-1) + \frac{G^2}{2(1-\alpha)^2\sigma\eta}, \forall \boldsymbol{\mu}.
\]
We have $ \| \boldsymbol{\mu}_t-\boldsymbol{\mu}_0\|_{\boldsymbol{\gamma}^2}^2\leq L^2$ according to the Cauchy-Schwarz' inequality.
The results follows.
Let  $M=\frac{L^2}{\eta} + \frac{G^2}{(1-\alpha)\sigma}\eta(T-1) + \frac{G^2}{2(1-\alpha)^2\sigma\eta}$. 
We now choose a proper $\boldsymbol{\mu}$, s.t. the complementary stackness can be further bounded.

For $\boldsymbol{\mu} = \hat{\boldsymbol{\mu}} + \theta$, where $\theta\in\mathbb{R}^{|P|}$ is non-negative to be determined later and $\hat{\boldsymbol{\mu}} = \argmax_{\boldsymbol{\mu}}-\boldsymbol{\mu}^{\top}(\sum_{i=1}^T\mathbf{A}^{\top}\mathbf{x}_t)/\lambda$. According to the constraint $\mathbf{e} = \sum_{i=1}^T\mathbf{A}^{\top}\mathbf{x}_t$, we have that
\[
    \sum_{t=1}^T(r(\mathbf{e}_t) + \boldsymbol{\mu}^{\top}\mathbf{e}_t\lambda) \leq r^*(-\hat{\boldsymbol{\mu}}) = r(\sum_{i=1}^T\mathbf{A}^{\top}\mathbf{x}_t) + \hat{\boldsymbol{\mu}}^T(\sum_{i=1}^T\mathbf{A}^{\top}\mathbf{x}_t)/\lambda.
\]
Note that in proof of Theorem 1, the feasible region $\mathcal{D}$ is recession cone, therefore, $\boldsymbol{\mu}\in\mathcal{D}$.

Therefore, we have 
\begin{equation}
\begin{aligned}
   \sum_{t=1}^{\tau}w_t(\boldsymbol{\mu}_t)
   &= \sum_{t=1}^Tw_t(\hat{\boldsymbol{\mu}}) - \sum_{t=\tau+1}^Tw_t(\hat{\boldsymbol{\mu}})  + \sum_{t=1}^{\tau}w_t(\theta) + M.\end{aligned}
\end{equation}

For each iteration $n$, we have 
\begin{align*}
    w_t{\bm{\mu}_t} - \hat{w}_t{\bm{\mu}_t} &= \bm{\mu_t}((\bm{M}-\bm{M})^{\top}\bm{x}_t) \leq \sum_{i,\bm{x}_{ti}=1} \triangle f_{u_t,i}^n L
\end{align*}

\textbf{Put them together:} 
For each batch $T$ at iteration $n$, we obtain that
\begin{equation}
    \begin{aligned}
        W_{OPT} &= \frac{\tau}{T}W_{OPT} + \frac{T-\tau}{T}W_{OPT} \\
        &\leq \tau W_{Dual}(\widetilde{\boldsymbol{\mu}_{\tau}}) + (T-\tau)(K+\lambda\bar{r}) \\
        & \leq \tau \hat{W}_{Dual}(\widetilde{\boldsymbol{\mu}_{\tau}}) + \sum_{t=1}^{\tau}\sum_{i, \bm{x}_{ti}}\triangle f_{u_t,i}^n + (T-\tau)(K+\lambda\bar{r})
    \end{aligned}
\end{equation}

Let's abbreviate $\triangle r(n) = \sum_{t=1}^{\tau}\sum_{i, \bm{x}_{ti}}\triangle f_{u_t,i}^n L$.
Therefore, combining Eq.~(10,12,13) the regret $\text{Regret}(n)$ of iteration $n$,  can be bounded as:
\begin{equation}
\begin{aligned}
    \text{Regret}(n) &= \mathbb{E}\left[W_{OPT}-\hat{W}\right]\\
    &\leq \mathbb{E}\left[W_{OPT}-\sum_{t=1}^{\tau}\left(\hat{g}(\mathbf{x}_t)/T - \lambda r(\bm{M}^{\top}\mathbf{x}_t/\bm{\gamma})\right)\right]\\
    &\leq \mathbb{E}\left[W_{OPT}-\tau \hat{W}_{Dual}(\widetilde{\boldsymbol{\mu}_t}) + \sum_{t=1}^{\tau}\hat{w}_t(\boldsymbol{\mu}_t) + \sum_{t=1}^T(\mathbf{e}_t-\bm{M}^{\top}\mathbf{x}_t)\right] \\
    &\leq \mathbb{E}\left[(T-\tau)(K+\lambda\bar{r})+\sum_{t=1}^Tw_t(\hat{\boldsymbol{\mu}}) + \sum_{t=1}^{\tau}w_t(\theta)\right] + M + \triangle r(n)\\
    &\leq (T-\tau)(K+\lambda\bar{r}+\lambda K) + \sum_{t=1}^{\tau}w_t(\theta) + M + \triangle r(n)\\
    &= T\text{Regret}(\pi^F)/N + \triangle r(n)
\end{aligned}
\end{equation}

Let $C = K+\lambda\bar{r}+\lambda K$, then setting the $\theta = {C}{\min_p \boldsymbol{\gamma}_p}\mathbf{u}_p$, where $\mathbf{u}_p$ is the p-th unit vector. We have
\[
    \sum_{t=1}^{\tau}w_t(\theta) = C/(\min_p \boldsymbol{\gamma}_p)  - C(T-\tau).
\]
Then the $\text{Regret}(n)\leq M + C/(\min_p \boldsymbol{\gamma}_p)$, when we set $\eta = O(T^{-1/2})$, the Regret($\pi^F$) is comparable with $O(T^{-1/2})$.

According to our algorithm in Algorithm of LTP-MMF, the user will interact with the system in $N/T$ iterations. Following the Lemma 11 in~\citet{abbasi2011improved}, the error $\text{Regret}(\pi)$ raised of accuracy module is bounded as 

\begin{align*}
    \text{Regret}(\pi) &= \sum_{n=1}^{N/T} \triangle r(n)
    \leq KT [\alpha_{N/T}(\rho_u+\kappa) + \beta_{N/T}(\rho_i+\kappa)],
\end{align*}
where 
\[
\kappa = \frac{(q+\epsilon_q)(1-(q+\epsilon_q)^{N/T})}{1-q-\epsilon_q},
\]
and 
\[
\rho_u = \sqrt{2d\frac{N}{T}\ln(1+\frac{N}{T\lambda_u d})},
\rho_i = \sqrt{2d\frac{N}{T}\ln(1+\frac{N}{T\lambda_i d})}.
\]
From the conclusion, we can see that the regret of accuracy module is comparable with $O(\sqrt{NT\ln\frac{N}{T}})$.


Finally, the total regret can be bounded as
\begin{equation}
    \begin{aligned}
        &\text{Regret}(\pi,\pi^F) = \sum_{n=1}^{N/T}\text{Regret}(n) = \text{Regret}(\pi^F)+\text{Regret}(\pi)L\\
    \end{aligned}
\end{equation}
Setting the learning rate as $\eta  = O(T^{-1/2})$, we can obtain a fairness regret Regret($\pi^F$) upper bound of order  $O(\frac{N}{\sqrt{T}})$. Overall, the long-term regret of LTP-MMF can be obtained of order $O(N\ln N)$.

\end{proof}
\bibliographystyle{ACM-Reference-Format}
\balance
\bibliography{ref}
    
\end{document}